\documentclass[journal=jacsat,keyval=apchd5,manuscript=article]{achemso}

\usepackage{chemformula} 
\usepackage[T1]{fontenc} 
\usepackage{siunitx}

\def\resp{\textit{resp. }}
\def\Ge{\Gamma_e/\Gamma_0}

\def\Geper{\Gamma_e^\perp/\Gamma_0}

\def\Gepar{\Gamma_e^\parallel/\Gamma_0}

\def\Gm{\Gamma_m/\Gamma_0}

\def\Gmper{\Gamma_m^\perp/\Gamma_0}
\def\GmperI{\Gamma_0/\Gamma_m^\perp}
\def\Gmpar{\Gamma_m^\parallel/\Gamma_0}

\def\r0{\textbf{r}_\text{0}}

\author{Yoann Brûlé}
\affiliation{ICB, Université Bourgogne-Franche Comté, CNRS, Dijon, France}
\email{yoann.brule@u-bourgogne.fr}
\author{Peter Wiecha}
\affiliation{LAAS, Université de Toulouse, CNRS, Toulouse, France}
\author{Aurélien Cuche}
\affiliation{CEMES-CNRS, Université de Toulouse, CNRS, Toulouse, France}
\author{Vincent Paillard}
\affiliation{CEMES-CNRS, Université de Toulouse, CNRS, Toulouse, France}
\author{Gérard Colas des Francs} 
\affiliation{ICB, Université Bourgogne-Franche Comté, CNRS, Dijon, France} \email{gcolas@u-bourgogne.fr}

\title[Purcell factor optimization]{Magnetic and electric Purcell factor control through geometry optimization of high index dielectric nanostructures}


\begin{document}

\begin{abstract}
We design planar silicon antennas for controlling the emission rate of magnetic or electric dipolar emitters.  Evolutionary algorithms coupled to the Green Dyadic Method lead to different optimized geometries which depend on the nature and orientation of the dipoles. We discuss the physical origin of the obtained configurations thanks to modal analysis but also emphasize the role of nanoscale design of the LDOS. We complete our study using finite element method and demonstrate an enhancement up to $2\times 10^3$ of the magnetic Purcell factor in europium ions. Our work brings together random optimizations to explore geometric parameters without constraint, a first order deterministic approach to understand the optimized designs and a modal analysis which clarifies the physical origin of the exaltation of the magnetic Purcell effect.
\end{abstract}

\section{Introduction}
The coupling of the magnetic part of light to atoms is much weaker than the electric one. Indeed, magnetic dipole transitions are $\alpha^2$ weaker than electric dipole transitions, where $\alpha \simeq 1/137$ is the fine-structure constant \cite{Berestetskii:1982}. Hence, the development of novel applications based on magnetic response in the optical regime, such as negative-index metamaterials \cite{Shalaev:2007} or efficient nanoantennas \cite{Bidault-Mivelle-Bonod:19}, requires to focus on the engineering of the magnetic Local Density Of States (LDOS) in order to enhance the magnetic contribution to light matter interaction. As predicted by E. M. Purcell more than 70 years ago for "nuclear magnetic moment at radio frequencies", the interaction of light and especially the spontaneous emission rate of solid state emitter can be drastically enhanced by its surrounding photonics environment which is well-known as the \textit{Purcell effect} \cite{Purcell:1946}. Placing a dipolar emitter into an optical micro-cavity or near a resonant nanostructure, it is possible to  control its emission rate. Until recently, interactions at a sub-wavelength scale were mainly focused on plasmonics, as noble metals nanostructures support Localized Surface Plasmon Resonances (LSPR) that are tunable by size, shape and constituent materials \cite{Novotny:2011,GCF-Barthes-Girard:2016}. However, despite impressive advances, severe limitations on the use of metals, such a high dissipation losses and poor compatibility with Complementary Metal Oxide Semiconductor (CMOS) technology, prevent them to be used in integrated devices. To overcome these limitations, replacing plasmonic resonators by high refractive index dielectric ones such as Silicon (Si) becomes interesting \cite{Kuznetsov:2016}. As the Si refractive index is above 3.5 and is associated to a very low extinction coefficient below its direct bandgap \cite{Baranov:2017}, it allows to obtain electric and magnetic Mie resonances in the visible to the near-infrared domain using nanostructures with sub-wavelength dimensions. Moreover, the high index dielectric contrast between the Si nanostructure and its low index environment (\textit{e.g.} $n \simeq 1.5$ for silica) ensures a high confinement and near-field intensity.  The use of Si also guarantees a fully compatible CMOS technology  for fabrication with large scalability and perfect reproducibility \cite{Kuznetsov:2016, Wiecha:2017, Staude:2017, Wood:2017}. Regarding the quantum emitters, rare-earth ions such as Europium ions (Eu$^{3+}$) are particularly relevant as they exhibit efficient electric and magnetic transitions in the visible domain \cite{Karaveli-Zia:2011,Aigouy:14,Kasperczyk:2015,Rabouw-Norris:16,Wiecha-Cuche:19,Bidault-Mivelle-Bonod:19,Chacon-GCF:2020,Majorel:2020}. Moreover, synthesis of sesquioxides (Y$_2$O$_3$, Gd$_2$O$_3$) thin films doped with such luminescent elements has become accessible \cite{Wiecha-Cuche:19} allowing to investigate the design of planar high index dielectric cavities.\\
In this context, numerical optimization is a particularly relevant tool. Since few decades, it has been largely applied to various domains of nanophotonics. While first attempts were focusing on the inverse design of optical coatings and mutilayered structures \cite{Dobrowolski:1988,Tikonravov:1996}, development of efficient algorithms associated to numerous flexible computational tools and improvement in computational power \cite{Moore:1995}, have allowed the design of various optical components with desired properties \cite{Molesky:2018} such as plasmonic \cite{Feichtner:2012, FeichtnerEO:2017, Wiecha:2018, Wiecha:2019} or dielectric \cite{Wiecha:2017,Bonod:2019} nanoantennas, compact broadband on-chip wavelength demultiplexer \cite{Pigott:2015} or plasmonic and dielectric metasurfaces \cite{Elsawy:2020}. To do so, various classes of algorithms have been used such as gradient-based methods \cite{Fu:2005}, Evolutionary Optimization (EO) techniques \cite{Goldberg:1988} and more recently deep learning approaches \cite{Liu:2018,Yao:2019,So:2020, Wiecha:2021}.\\

In this work, we apply a subset of EO algorithms called Differential Evolution (DE) \cite{Storn:1997} in order to optimize the geometrical design of planar Si dielectric antennas coupled to Gd$_2$O$_3$ nanostructures doped with Eu$^{3+}$ emitters (Gd$_2$O$_3$:Eu$^{3+}$) for maximizing or minimizing the decay rate enhancements (\textit{i.e.} the Purcell factor) of their magnetic and electric dipolar transitions. We deeply investigate the magnetic Purcell factor to overcome the extremely weak magnetic light-matter interaction. We also discuss the electric Purcell factor as a benchmark since it has been widely explored in the literature.  To go beyond the "black box" optimization, we  interprete the physical origin of the obtained designs thanks to modal analysis but also in relation to the deterministic design approach recently developped by Mignuzzi \textit{et al.} in \cite{Mignuzzi:2019}. Finally, having determined the optimal shapes for enhancing or inhibiting the magnetic dipolar emission, we investigate the maximum and minimum achievable magnetic Purcell factor by finite element method simulations, better adapted to the cylindrical shapes suggested by EO optimization.
\section{Green Dyadic Method and differential evolution algorithm}
\label{section:Conv}
EO algorithms are inspired from the evolution theory \cite{Goldberg:1988}. These bio-inspired algorithms can be classified as stochastic ones as they iteratively use random mutations of an initial random population of candidates in order to make it evolves towards one population whose individuals present the best predefined characteristics. These characteristics can be implemented by one (single objective \cite{Islam:2012}) or several (multi-objective \cite{Wiecha:2017,Deb:2001}) fitness functions to maximize (or minimize). These algorithms aim to find the global optimal solution of problems with large and complex dimensionality. They can indeed be seen as global optimization techniques.
They have been recently applied to the better understanding of natural design of photonic architectures that has been optimized through natural evolution \cite{Barry:2020}. Regarding the optimization of nanoantennas geometry, EO algorithms have been applied to maximize the near-field intensity, Purcell factor and directivity considering dielectric \cite{Bonod:2019,Gondarenko:2006} or plasmonic nanostructures \cite{Feichtner:2012,FeichtnerEO:2017,Wiecha:2018,Wiecha:2019}.\\

In the following, we are presenting the results obtained by applying a specific type of EO algorithms, Differential Evolution \cite{Storn:1997} in order to optimize either the electric or magnetic decay rates. The emitters are europium ions doping Gd$_2$O$_3$ matrix since they present both  electric (ED) and magnetic (MD) dipolar transitions at wavelengths $\lambda_e = 610$ nm and $\lambda_m = 590$ nm, respectively. We investigate the enhancement of their emission by coupling them to planar silicon nanostructures of arbitrary shape. We have used the Green Dyadic Method (GDM) and specifically the python toolkit pyGDM \cite{pyGDM:2018,pyGDM:2021} to perform decay-rate calculation inside multi-material nanostructures \cite{Wiecha:2018}. The GDM is based on a volumic discretization of the nanostructures. In order to limit computational resources and consider realistic shapes, the cubic mesh size has been limited to 20 nm, slightly above the resolution of standard e-beam lithography. It is worth mentionning that comparison with Mie analytical model has demonstrated that the decay is strongly sensitive to the meshing \cite{pyGDM:2021}. We attribute this mainly to the strong sensitivity of morphological resonances to the object shape. For the current work we estimate the error of the order of 5\% for the electric Purcell factor but up to 30-50\% for magnetic Purcell factor. 
Since we will observe that the arbitrarily optimized structures also rely on resonances, we will complete this study in a second step, considering finite element method, better adapted to describe rounded object. \\

\begin{figure}[!h]
\centering
\includegraphics[width=0.49\columnwidth]{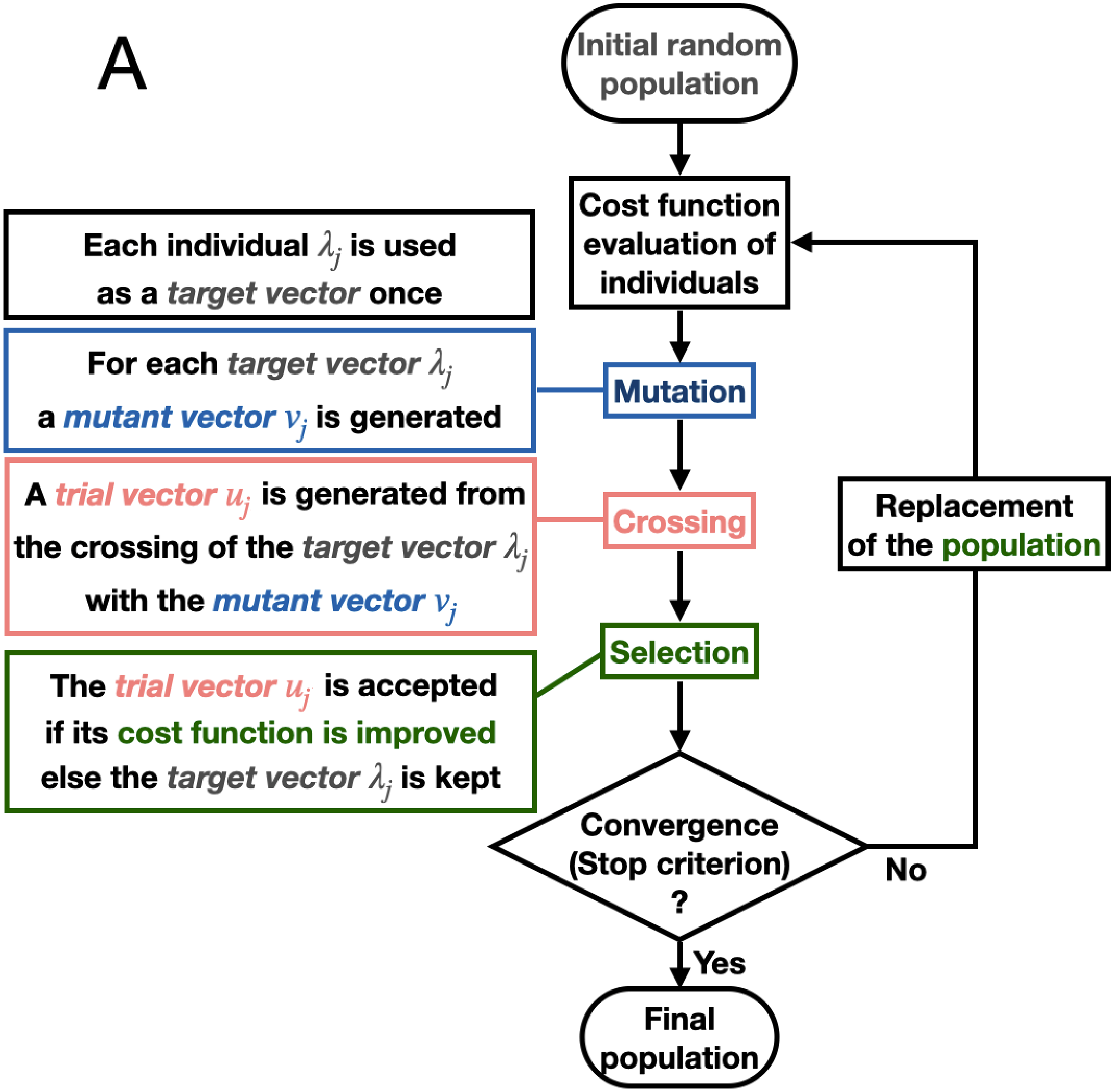}
\raisebox{0.65\height}{\includegraphics[width=0.5\columnwidth]{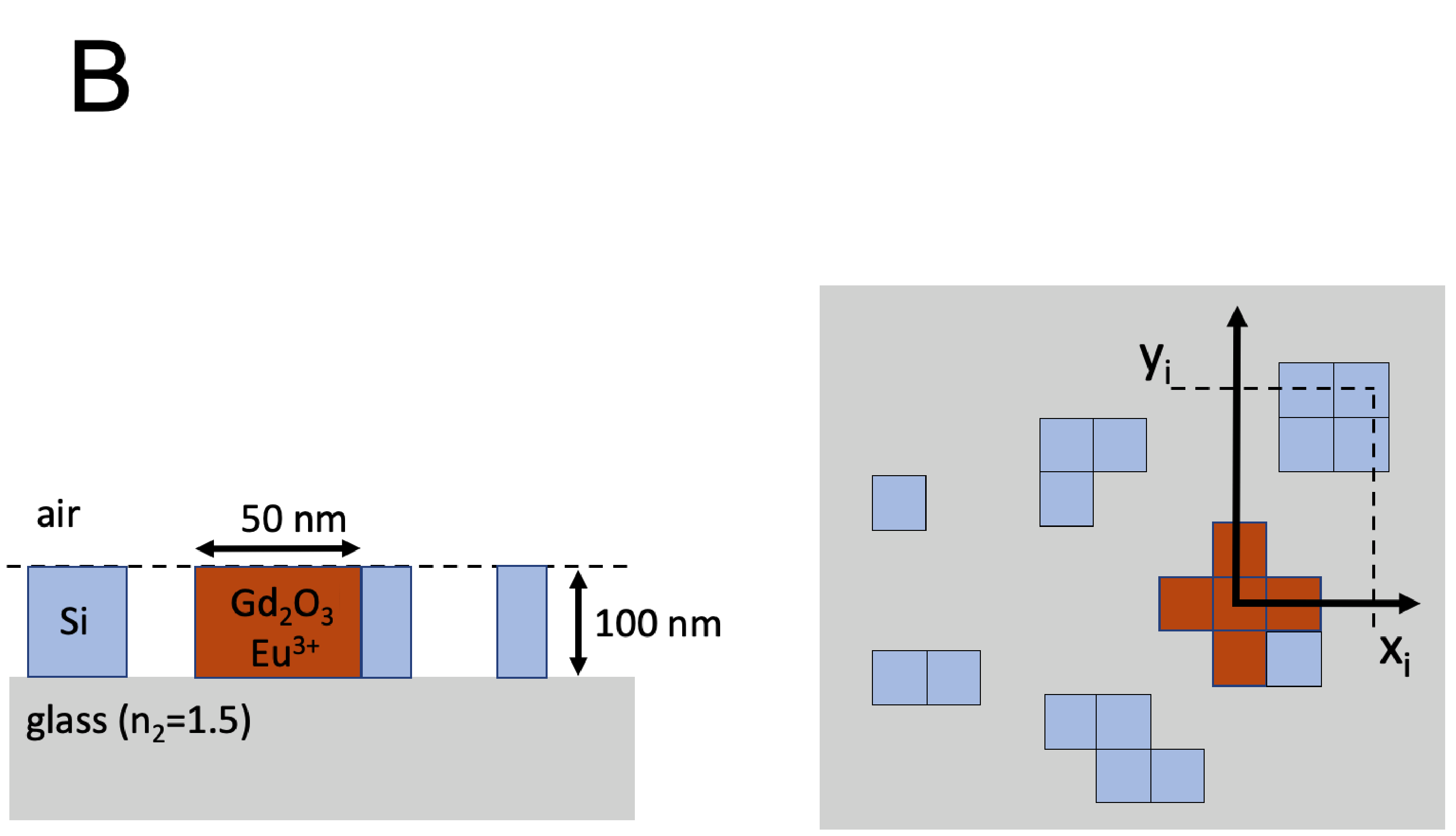}}
\caption{A) Algorithm of the differential evolution cycle for geometrical optimization. B) Scheme of the structure to be optimized. $N$ Si nanopillars (20x20x100 nm$^3$)  are free to evolve in a 1.68 $\times$ 1.68 $\mu$m$^2$ area.}
\label{fig:DE}
\end{figure}

The EO algorithm and the sample to be optimized are depicted in Fig.~\ref{fig:DE}. The core consists in a Gd$_2$O$_3$:Eu$^{3+}$ matrix of 100 nm height and 50 nm diameter (optical index $n_1 = 1.8$). We are optimizing the planar nanostructured Si environment which is constituted of $N = $ 300, 400, 500, and 600 Si nanopillars (each 20 nm $\times$ 20 nm $\times$ 100 nm) lying on a SiO$_2$ substrate ($n_2 = 1.5$) in an area limited to 1.68 $\times$ 1.68 $\mu$m$^2$. The surrounding medium is air. Refractive index of the silicon $n_{Si}$ has been taken from \cite{Palik:1997}. 
In reality, a Gd$_2$O$_3$:Eu$^{3+}$ core would contain a lot of randomly oriented emitters. However, for the sake of simplicity and understanding, we have limited the study to a single emitter with defined orientation. Thus, we have fixed the optimization goal to find the Si nanostructure that maximizes or minimizes the magnetic (\resp electric) decay rate enhancement $\Gm$ (\resp $\Ge$) at wavelength of $\lambda_m$ (\resp $\lambda_e$) for an emitter situated at the center of the core $\r0 = [0, 0, 50]$ nm and oriented either along or perpendicular to the substrate. 
A population of 64 individuals $\lambda_j$ ($j = 1$ to 64) is evolving. Each of which is a set of $(x_i,y_i)$ positions ($i = 1$ to $N$) chosen among a 20 $\times$ 20 nm$^2$ discretized grid of the 1.68 $\times$ 1.68 $\mu$m$^2$ plane, core emitter positions excluded (\textit{i.e.} $ 85^2 - 5 = 7220$ possible positions). To converge, EO algorithms apply the following process. The fitness function (Purcell factor) of each individual of the current generation is evaluated. A mutation and a crossover is performed to form trial individuals. The trial individuals fitness functions are then re-evaluated and a selection is operated in order to form a new generation with better characteristics. The process is then repeated iteratively (see Fig.~\ref{fig:DE}A). We have used the self adaptive "jDE" implementation of  differential evolution for single objective problem \cite{Islam:2012} available in the Python "pyGMO" interface associated to the "paGMO" library \cite{Biscani:2020} and to the Python toolkit pyGDM \cite{Wiecha:2017,pyGDM:2018,pyGDM:2021}.  
We are using the default "/rand/1/exp" parameter for the mutation variant of the jDE algorithm and its default configuration for the auto-adaptation scheme.

The next section presents the results of the optimization for the magnetic (\resp electric) dipolar emissions. For the sake of clarity, we discuss in the main text two representative configurations leading to the highest Purcell factor: namely out of plane MD and in-plane ED. The others configurations (in-plane MD and out of plane ED) lead to similar designs and are presented in the Supplement 1, section 1.

\section{Geometrical optimizations}
\subsection{Out of plane magnetic Purcell factor}
\label{section:EOMagZ}
\begin{figure}[!h]
\centering
\includegraphics[width=0.66\columnwidth]{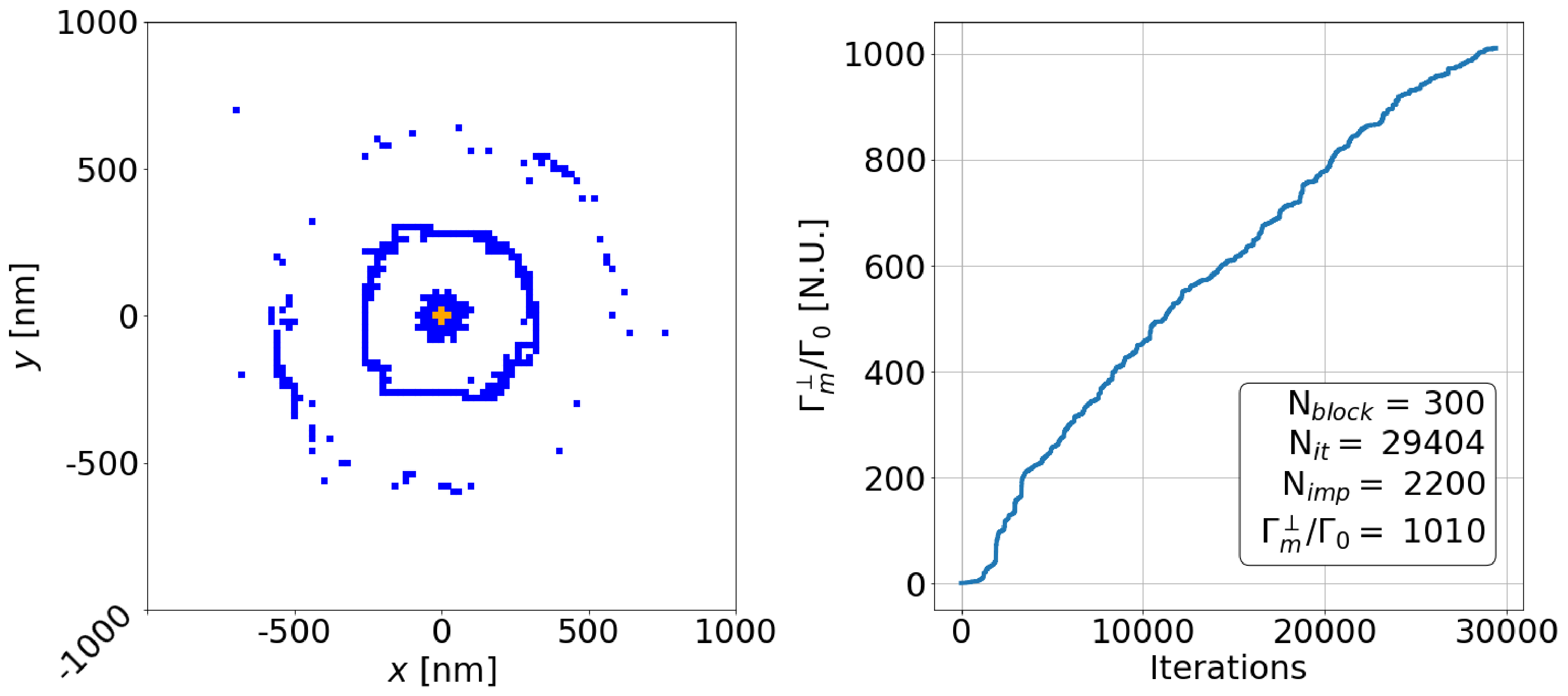}
\raisebox{1cm}{\includegraphics[width=0.33\columnwidth]{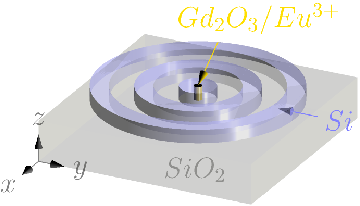}}
\caption{Left : $XY$-plane projection of the optimized structure (orange : fixed core emitter, blue : Si nanopillars), Center : Evolution of the magnetic Purcell factor $\Gmper$ through the optimization iterations, Right : Scheme of the configuration extrapolated from the optimized nanostructure.}
\label{fig:OptMagZ}
\end{figure}

In Fig.~\ref{fig:OptMagZ}, we present the results of the optimization for the decay rate enhancement of the magnetic dipolar emission of Eu$^{3+}$ at $\lambda_m=590$ nm and for an out of plane dipole.
We only show the results of the optimization for $ N = 300 $ Si blocks. Numerical simulations were repeatedly reproduced with differents numbers of Si blocks and all lead to similar shapes as well as optimized Purcell factors (see section 1.A in the Supplement 1).\\

The optimization has been stopped after $N_{it} = 29404$ iterations with $N_{imp} = 2200$ improvements of the fitness function ($\Gmper$) of the best candidate among the full 64 individuals of the population. It converges towards a regular structure that consists in a Si shell of diameter $\simeq 200$ nm surrounding the Gd$_2$O$_3$:Eu$^{3+}$ core and concentric Si rings of width $\simeq 100$ nm and period $\simeq 300$ nm.
The associated enhancement of the magnetic decay rate is evaluated to $\Gmper=1010$. 
We can presume that an optimization on a larger area with more Si blocks would converge toward a cylindrical grating.\\

\begin{figure}[!h]
\centering
\includegraphics[width=0.66\columnwidth]{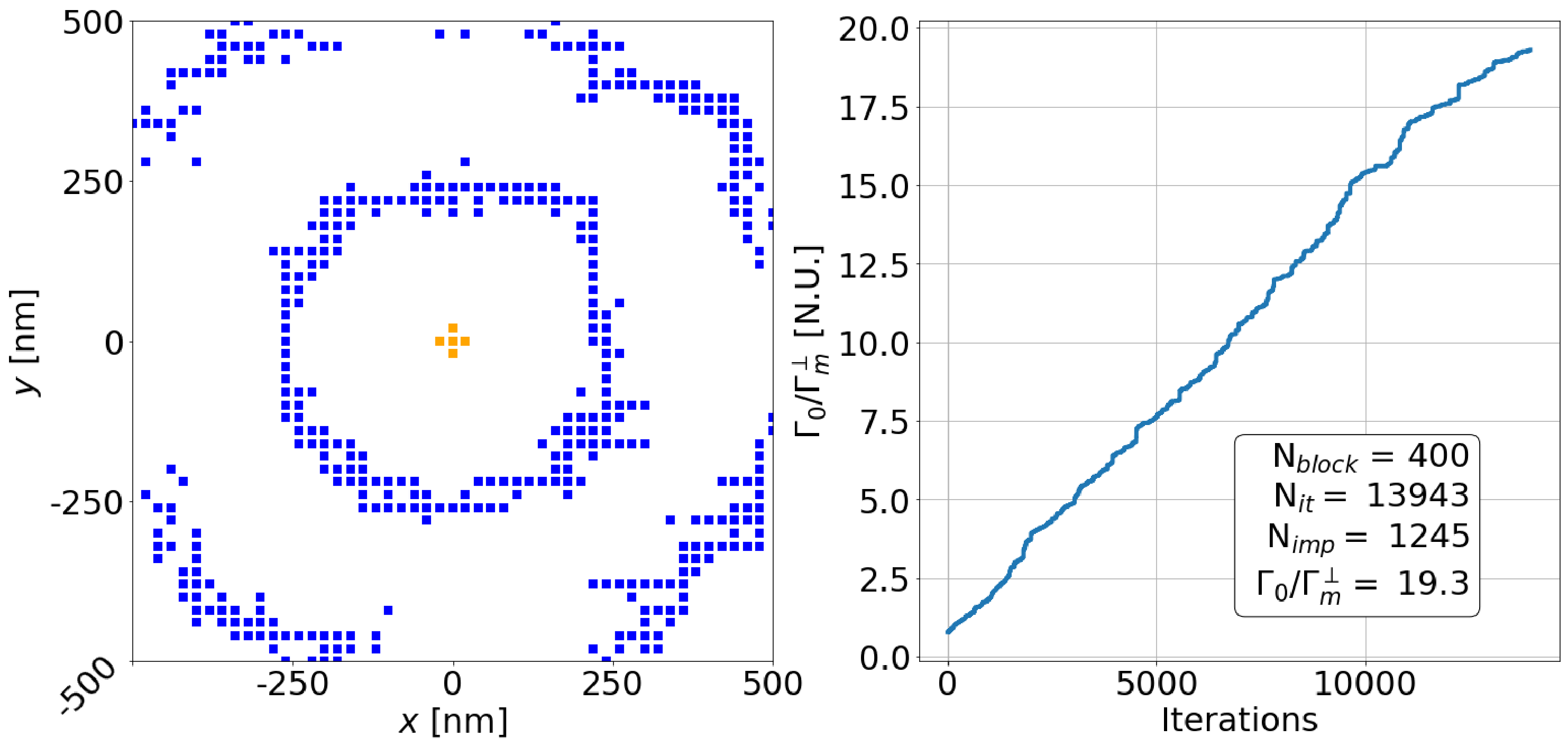}
\raisebox{1cm}{\includegraphics[width=0.33\columnwidth]{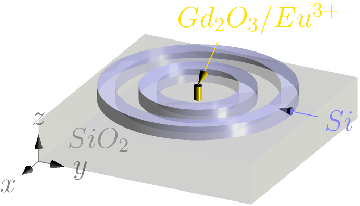}}
\caption{Left : $XY$-plane projection of the optimized structure (orange : fixed core emitter, blue : Si nanopillars), Center : Evolution of the magnetic decay rate inhibition $\GmperI$ through the optimization iterations, Right : Scheme of the configuration extrapolated from the optimized nanostructure. The area of optimization is limited to a 500$\times$500 nm$^2$ plane and the number of block is $N$ = 400.}
\label{fig:OptMagZInhib}
\end{figure}

Reversely, we can minimize the MD transition at $\lambda_e=590$ nm, see Fig.~\ref{fig:OptMagZInhib}. It converges towards a regular structure that consists in concentric Si rings of width $\simeq 80$ nm and period $\simeq 250$ nm without shell around the core emitter. The associated inhibition of the magnetic decay rate is evaluated to approximately $\Gmper \simeq 1/19$.  We can observe that the optimized structure for the inhibition is complementary to the one of the exaltation.\\

\subsection{In-plane electric Purcell factor}
\begin{figure}[!h]
\centering
\includegraphics[width=0.66\columnwidth]{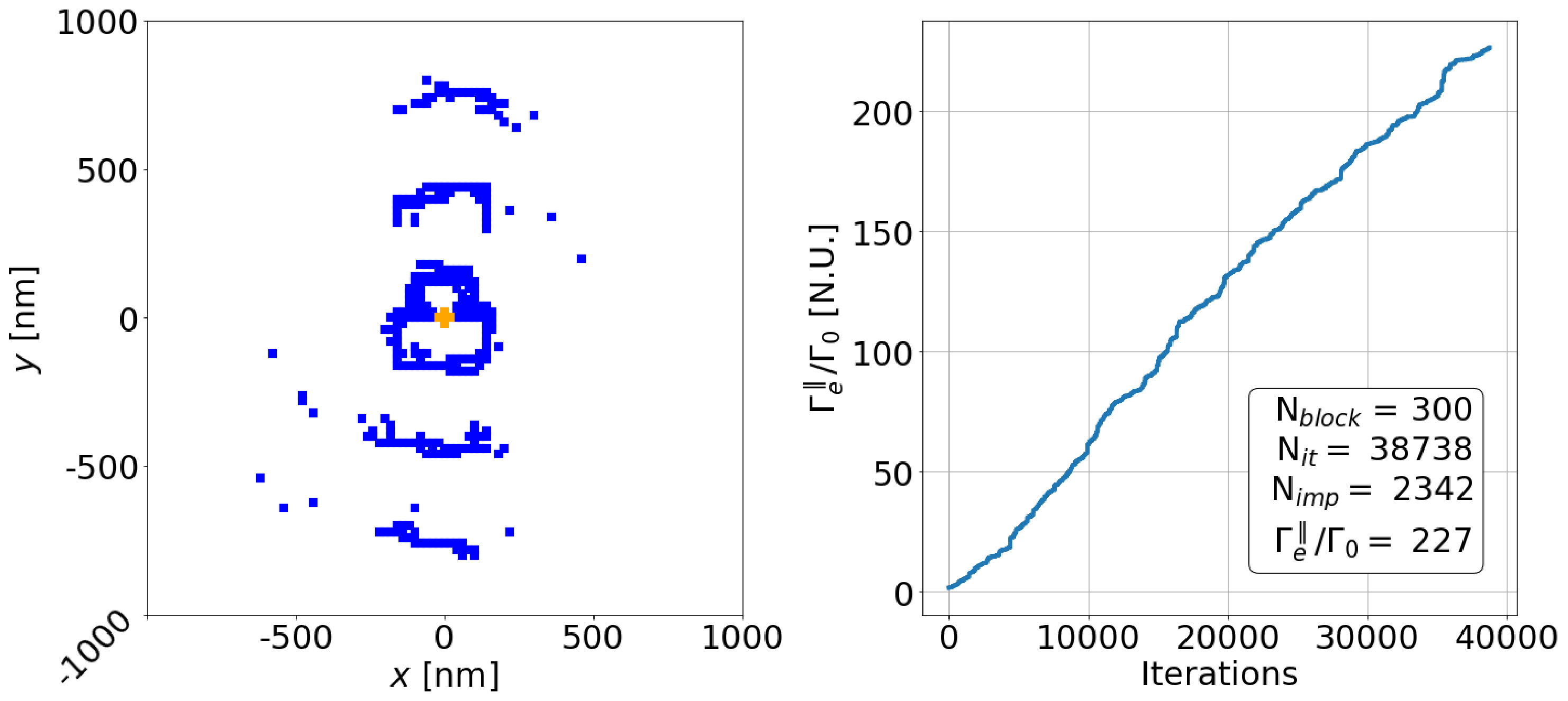}
{\includegraphics[width=0.3\columnwidth]{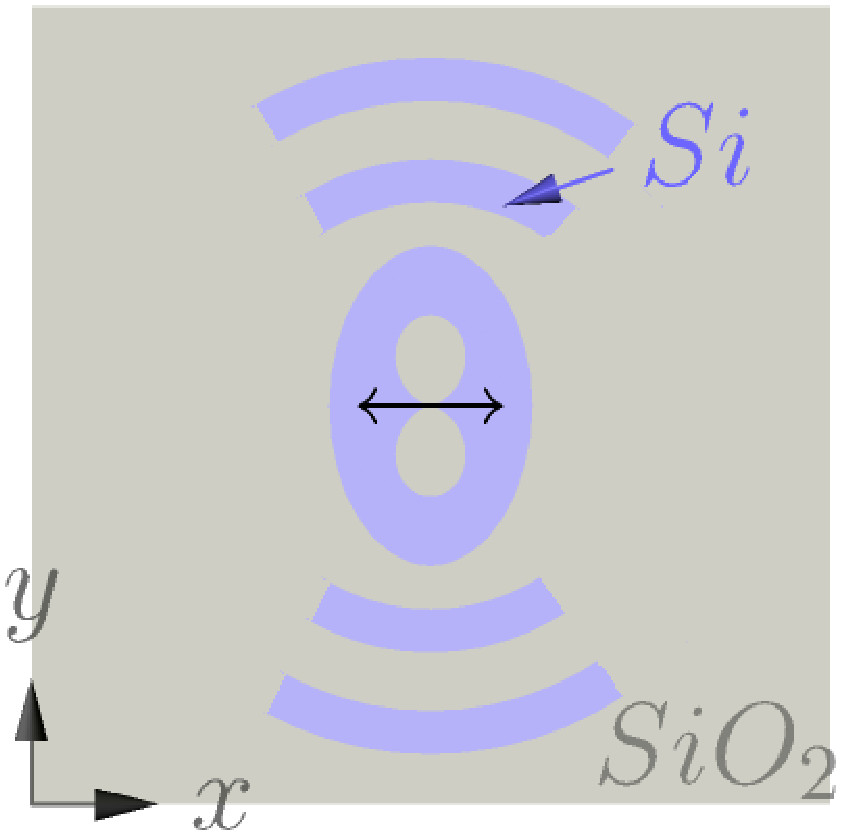}}
\caption{Left : $XY$-plane projection of the optimized structure (orange : fixed core emitter, blue : Si nanopillars), Center : Evolution of the electric Purcell factor $\Gepar$ through the optimization iterations, Right : Scheme of the configuration extrapolated from the optimized nanostructure. The ED source is along the $x$-axis (black arrow).}
\label{fig:OptElecX}
\end{figure}

\begin{figure}[!h]
\centering
\includegraphics[width=0.4\columnwidth]{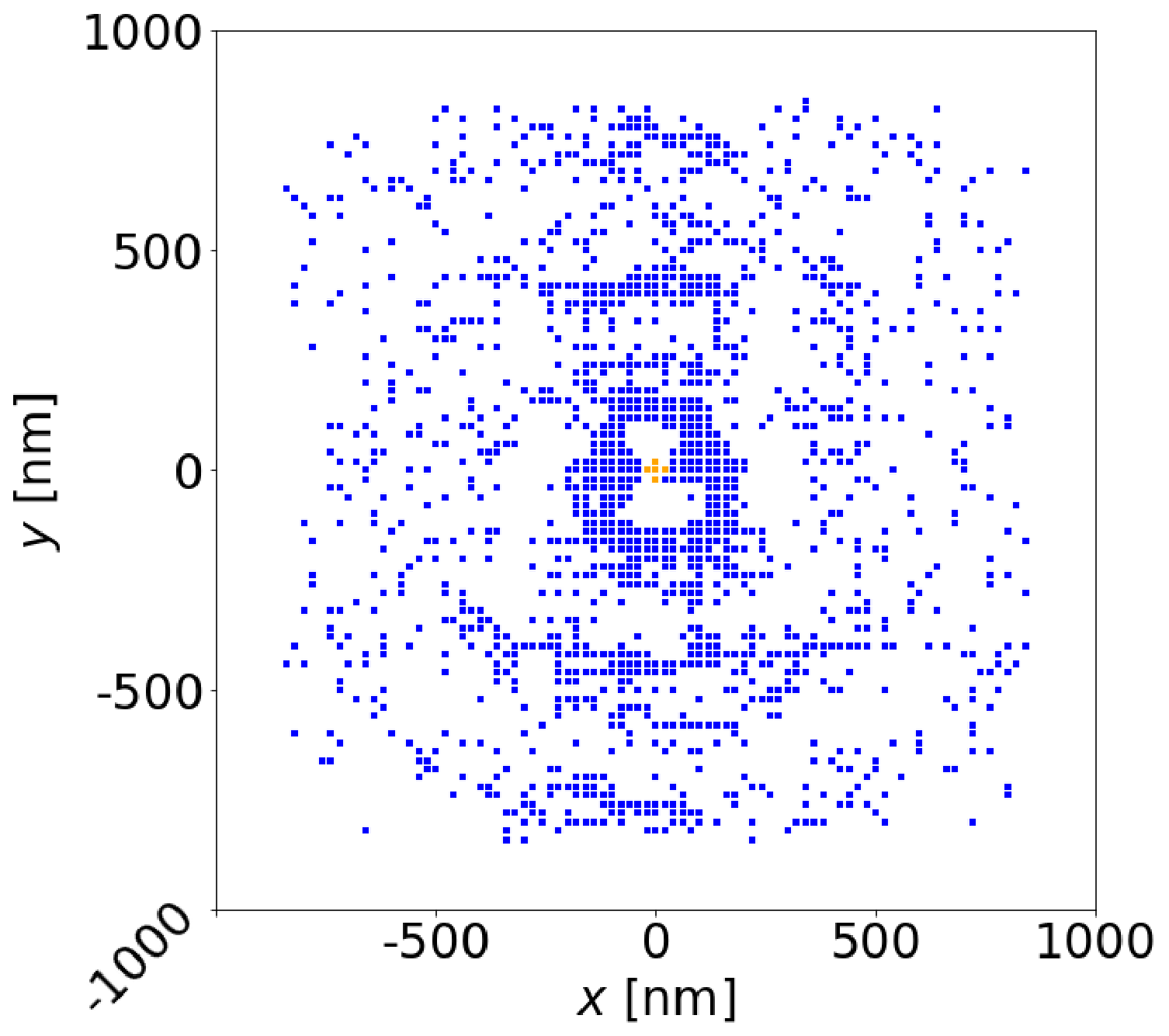}
\caption{$XY$-plane projection of the structure arising from the superposition of four independant EO evolving $N=$ 300, 400, 500 or 600 Si blocks (orange : fixed core emitter, blue : Si nanopillars).}
\label{fig:OptElecXSup}
\end{figure}

For comparison, we present in Fig.~\ref{fig:OptElecX} the optimization of the in-plane electric Purcell factor at $\lambda_e = 610$ nm, considering $ N = 300 $ Si blocks. 
The optimized structure leads to an in-plane electric Purcell factor $\Gepar \simeq 227$.\\
Regularity is emerging one more time. A bowtie aperture antenna is formed and circular rings also appear. A qualitative scheme extrapolated from the optimized structure is represented on the right part of Fig.~\ref{fig:OptElecX} (see also in section Discussions).  Additional simulations with different numbers of Si blocks lead to similar results. Figure~\ref{fig:OptElecXSup} presents the superposition of the structures obtained optimizing with $N=$ 300, 400, 500 and 600 Si nanopillars. This figure does not result from an optimization but emphasizes that all optimized structures present similar features, namely a bowtie aperture antenna and circular rings.


\section{Discussion}
\label{subsection:Discussions}
To summarize, we distinguish different behaviours in the near and far field zones of the dipolar emitter.  In the far-field, we observe circular gratings for all the configurations. 
In case of in-plane ED (Fig. \ref{fig:OptElecX}) or MD  (see Fig. S3 for N=300 in the Supplement 1), the Si grating seems not fully circular but rather presents two lobes perpendicularly to the dipole orientation. Since dipolar emission is mainly along its axis in the near-field and perpendicular to its axis in the far-field, the presence (or not) of matter along the dipole axis in the far-field zone poorly modifies the dipole emission. Therefore, a simplified structure with full rings would present similar behaviours. As far as the near-field zone is concerned, we observe strong differences for ED and MD emitters, with the presence or not of a Si core, that depends on both the dipole nature and orientation. In case of in-plane ED, a complex bowtie aperture nanoantenna is retrieved. This bowtie shape reminds notably the nanostructures obtained in \cite{Gondarenko:2006, FeichtnerMode:2017, Mignuzzi:2019,Xie:2017}. In ref. \cite{Gondarenko:2006}, an EO algorithm was applied to an analog situation, that is improving the confinement of an in-plane electric field. Hence similar optimized design is retrieved since the Purcell factor is inversely proportional to the mode volume. In ref. \cite{FeichtnerMode:2017} the authors introduced the concept of mode matching to optimize the coupling of an electric emitter to a plasmonic nanoantenna but obtained similar features, revealing the generality of the achieved configurations. In case of MD, we observe a nanodisk that surrounds directly the emitters. Such geometry has already been spotted as good candidates for maximizing the magnetic Purcell factor \cite{Li:2017,Yang:2019,Aslan:2022}, our work confirms them to be close to the optimal. We also emphasize that these previous works obtain Purcell factor in the range 100-5000 but with a strong sensitivity to the presence of the substrate and the  diameter of the doped core.\\

\subsection{Material local contribution to the decay rates}
Aiming to access physical understanding of the achieved optimized structures, we use the approach of Mignuzzi and coworkers on the "nanoscale design of the LDOS" \cite{Mignuzzi:2019}. To this purpose, they have derived a volume integral for the expression of the ED decay rate enhancement so that they can assess {\it locally} the effect of matter on the electric decay rate, namely enhancement or inhibition. This reveals where to remove materials to strenghten the enhancement effect solely. They obtain that the decay rate associated to an electric dipole $\mathbf{d}$ located at $\mathbf{r}_d$ is expressed as 
\begin{eqnarray}
\frac{\Gamma_e}{\Gamma_0}&=&1+\frac{6\pi\epsilon_0}{k_0^3\vert d\vert^2}Im\left\{ \mathbf{d} \cdot \mathbf{E}_s(\mathbf{r}_d)\right\}\\
&=&1+\frac{6\pi}{k_0^3\vert d\vert^2}\int d^3{\mathbf r} ~ \left(\epsilon_r({\mathbf r})-1\right) Im\left[ f_E(\mathbf{r})\right] \;, \text{with}\\
&& f_E(\mathbf{r})=\mathbf{d} \cdot {\mathbf G}_0({\mathbf r}_0,{\mathbf r}) \cdot {\mathbf G}({\mathbf r},{\mathbf r}_0) \cdot \mathbf{d}
\end{eqnarray}
where $\mathbf{E}_s$ is the electric field scattered by the dipolar source in the complex environment, $k_0$ the free-space wavenumber. In the two last lines, we introduced the free-space electric Green's tensor ${\mathbf G}_0$ and the Green's tensor associated to the complex environment ${\mathbf G}$.
Mignuzzi and coworkers propose to remove material everywhere where $Im\left[ f_E(\mathbf{r}) \right] < 0$ since it induces decreasing of the decay rate. By an iterative procedure they finally obtained {\it deterministically }complex geometries presenting similar feature as "blackbox" optimization. 

In section 2 of the Supplement 1, we derive an analog expression for the magnetic dipole emission 
\begin{eqnarray}
 \frac{\Gamma_m}{\Gamma_0}&=&1+\frac{6\pi}{\mu_0k^3\vert m\vert^2}Im\left\{ \mathbf{m} \cdot \mathbf{B}_s(\mathbf{r}_d)\right\}\\
 &=&1+\frac{6\pi}{k^3\vert m\vert^2}\int d^3{\mathbf r} ~ \left(\epsilon_r({\mathbf r})-1\right) Im\left[ f_H(\mathbf{r})\right] \\
&& f_H(\mathbf{r})=\mathbf{m} \cdot {\mathbf G}_0^{HE}({\mathbf r}_0,{\mathbf r}) \cdot {\mathbf G}^{EH}({\mathbf r},{\mathbf r}_0) \cdot \mathbf{m}
\end{eqnarray}
where  the mixed Green's tensor ${\mathbf G}^{EH}$ gives the electric field scattered by a magnetic dipole in the complex environment. ${\mathbf G}^{HE}_0$ refers to the magnetic field scattered by an ED in free--space. 

\begin{figure}[h!]
\includegraphics[width=0.45\columnwidth]{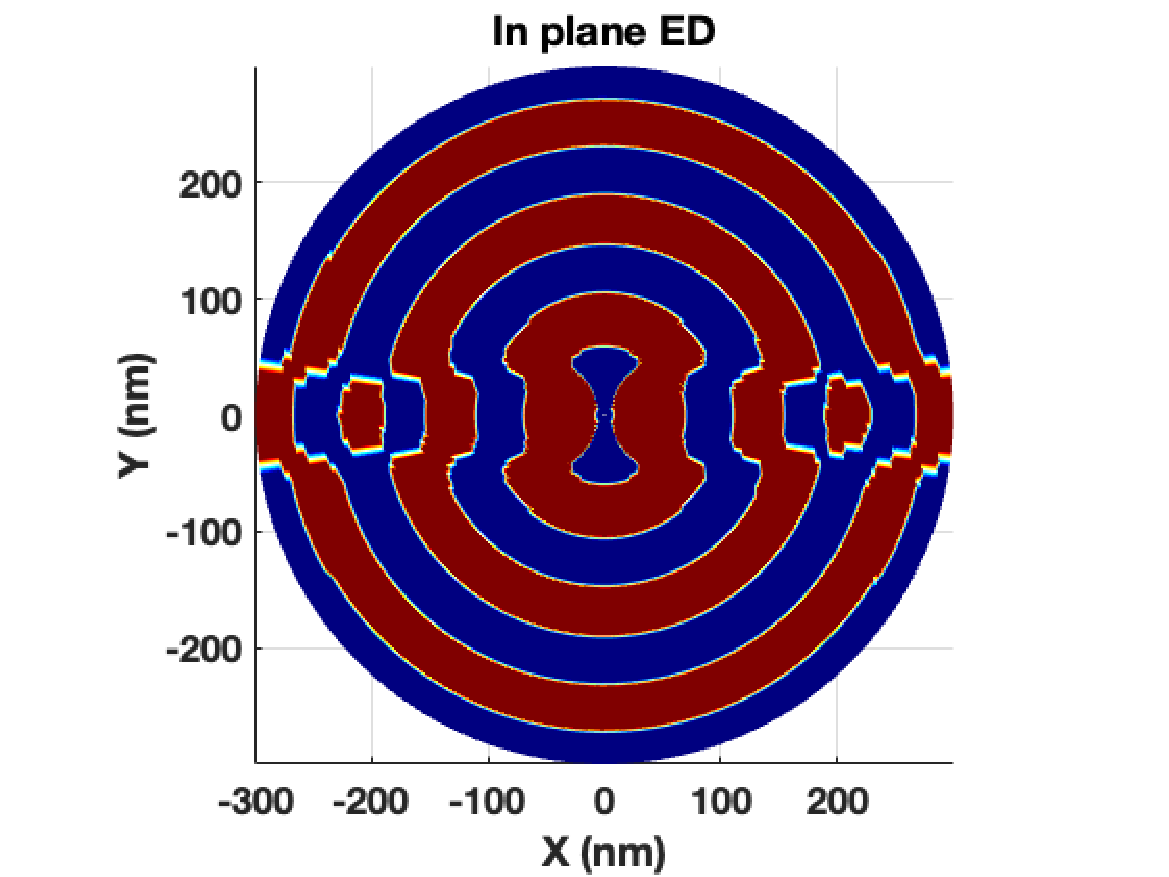}
\includegraphics[width=0.45\columnwidth]{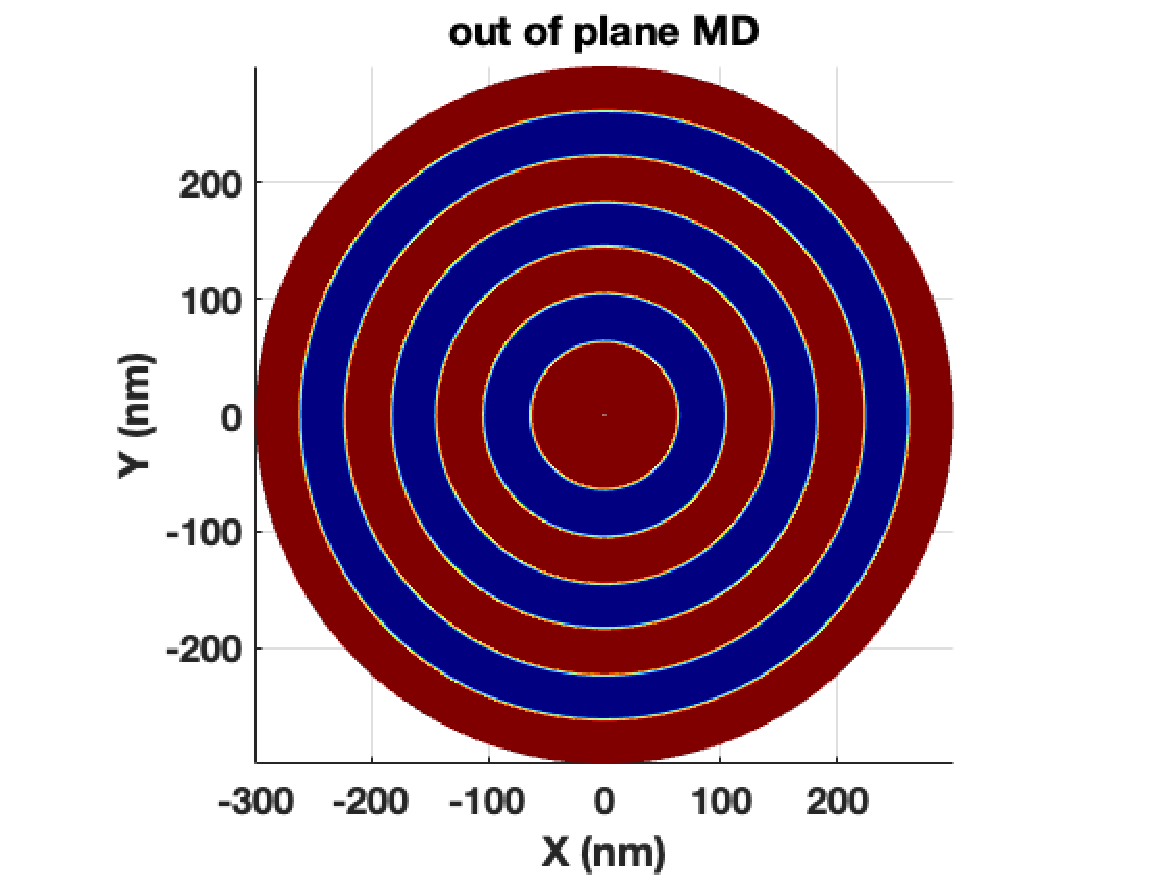}
	\caption{Sign of the Im(f) for in-plane (along $x$, $\lambda_e=610$ nm) electric and out of plane (along $z$, $\lambda_m=590$ nm) magnetic dipoles. The dipole is in the middle of a Si layer (100 nm) sandwiched between a glass substrate and air superstrate. Blue and red areas refer to negative and positive signs, respectively. 
	\label{fig:SapienzaED-MD}}
\end{figure}
We consider the  dipolar emission in the silica/Si(100nm)/air slab for which Green's tensors are analytical. We plot on figure \ref{fig:SapienzaED-MD} the sign of the factor $Im(f)$. Blue areas with $Im(f)<0$ reveal where materials has to be removed to enhance the decay rate. Better designs should be obtained from an iterative procedure \cite{Mignuzzi:2019} but we use this approach solely to qualitatively understand the shapes obtained by EO approach. For an electric dipole parallel to the interfaces, we recover the bowtie aperture antenna configuration with an additional circular grating.  For out of plane MD, we observe a circular symmetry, again in agreement with our EO optimization, with notably the presence of a high-index dielectric core.
We observe qualitative agreement between the shape suggested by removing material where it reduces the LDOS. However, we didn't pursue an iterative process to recalculate the positive or negative local contributions of material to the LDOS at the dipole position. Therefore, no quantitative agreement could be achieved, particularly on the ring periodicity since removing material will strongly modify the effective wavelength in the nanostructured medium. Nevertheless, the approach proposed by Mignuzzi and coworkers clearly reveals the physical origin of the EO optimized structures design. In addition, we emphasize that ED and MD emissions are fully analog in free-space. Different reflexions at the silica/Si and Si/air interfaces for ED or MD emission is at the origin of the different optimized designs.\\

Reversely, removing material everywhere where $Im(f)>0$ (red areas) will minimize the Purcell factor, in agreement with the EO simulation presented in Fig. \ref{fig:OptMagZInhib}. This also demonstrates the efficiency of the EO geometrical optimization that can be extended safely to other criteria for which no deterministic approach exists.

\subsection{Modal analysis}
\begin{figure}[!h]
\centering
\includegraphics[width=0.95\columnwidth]{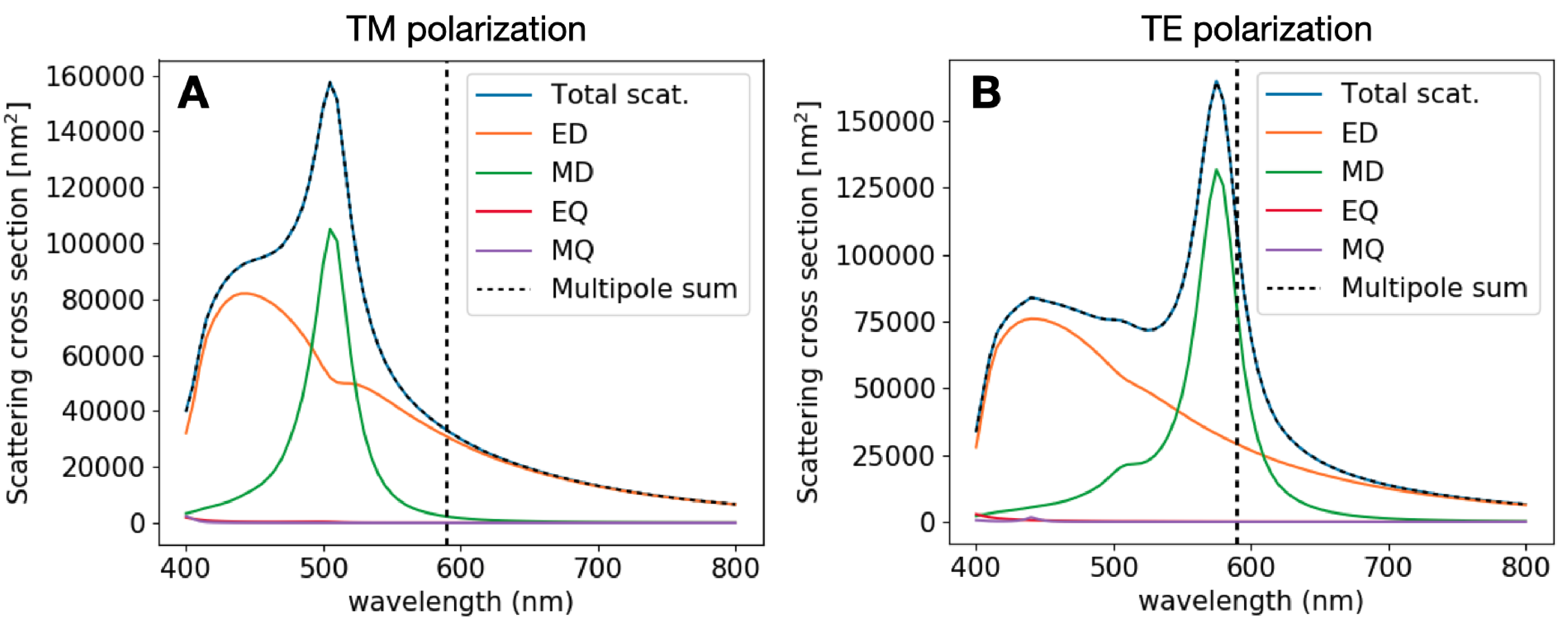}
\caption{Spectral decomposition of the scattering cross section of the Si core when irradiated from the substate by a plane wave with oblique incidence ($\SI{40}{\degree}$) with (A) TM and (B) TE-polarization. The vertical dashed black lines stand for the optimized wavelength of the Eu$^{3+}$ MD emission $\lambda_m = 590$ nm.}
\label{fig:CoreSCS}
\end{figure}

For a better understanding of the MD decay rate enhancement mechanisms, we have performed a multipole expansion of the optical response \cite{pyGDM:2021,Alaee:2018} on the cylindrical Gd$_2$O$_3$:Eu$^{3+}$ core/Si shell nanostructure found in Section \ref{section:EOMagZ}.
We first consider plane wave excitation of the Gd$_2$O$_3$:Eu$^{3+}$ core (50 nm diameter) with a Si shell (44 nm width). We have used  hexagonal meshing of 9 nm for better description of the circular shape inferred from both EO arbitrary optimization and nanoscale design of the LDOS. The considered dimensions of the core correspond to the maximum decay rate enhancement for this shape ($\Gmper \simeq 256$, see also the next section). In figure \ref{fig:CoreSCS}, we plot the
scattering cross section for plane wave illumination from the substrate. We have considered $\SI{40}{\degree}$ oblique incidence to access both in-plane and out of plane modes of the structure. 
We observe two MD and one ED resonances. We associate the MD resonances at 510 and 580 nm to in-plane and out-of plane MD modes, respectively.

\begin{figure}[!h]
\centering
\includegraphics[width=0.95\columnwidth]{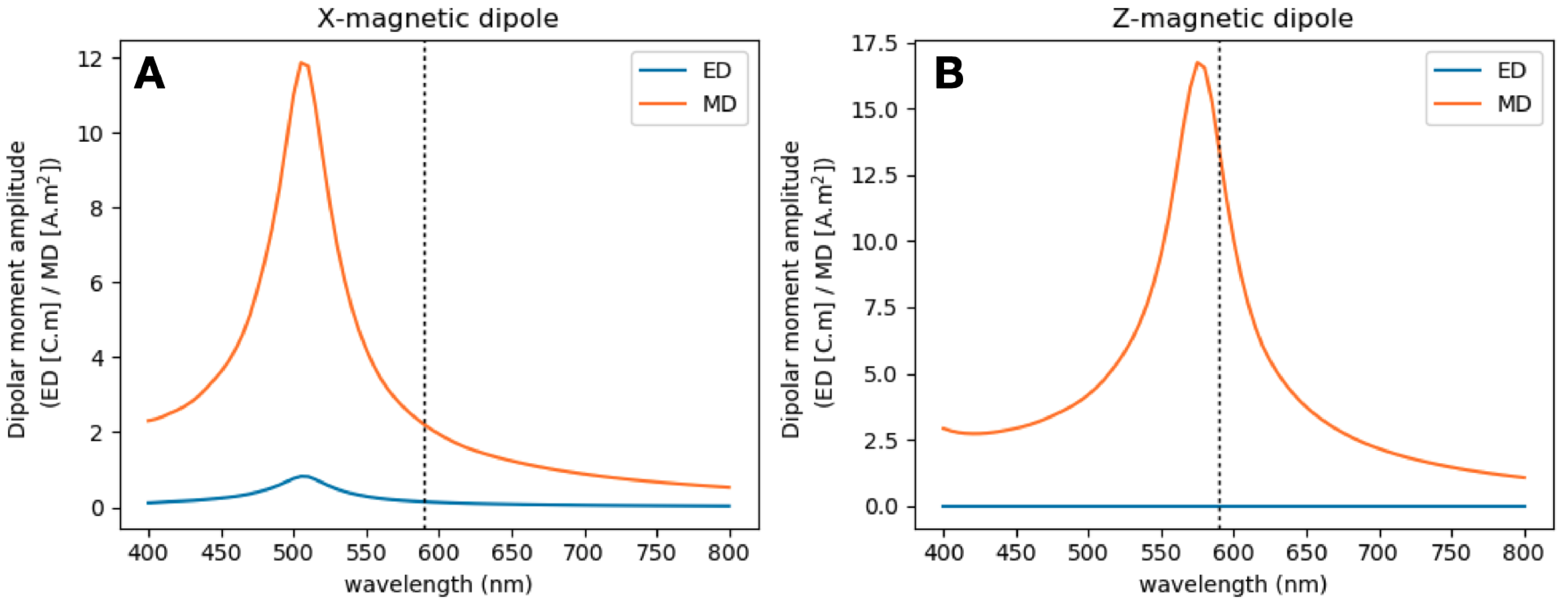}
\caption{Spectral decomposition of the internal field response of the core structure to an MD excitation with a dipolar moment amplitude of 1 A.m$^{2}$ (A) oriented along the $x$-axis (in-plane) and (B) along the $z$-axis (out of plane). The vertical dashed black lines stand for the optimized wavelength of emission $\lambda_m = 590$ nm.}
\label{fig:CoreDipolarDecomp}
\end{figure}

We have then calculated the multipolar decomposition of the scattered field of the same structure but excited with an in- and out-of-plane MD, placed at its center, see Fig. \ref{fig:CoreDipolarDecomp}.
On the one hand, the in-plane magnetic dipole couples mainly to the in-plane MD mode of the structure at 510 nm plus a small ED contribution. The amplitude of the magnetic dipole induced in the nanostructure is 12 times higher than the MD source. On the other hand, the out of plane magnetic dipole couples to the out of plane MD mode of the structure at 580 nm. The induced MD amplitude is enhanced by a factor 17.\\


\subsection{Maximum and minimum achievable magnetic Purcell factor}
\label{sec:FEMoptMagEnh}
\begin{figure}[!h]
\centering
\raisebox{0.9\height}{\includegraphics[width=0.45\columnwidth]{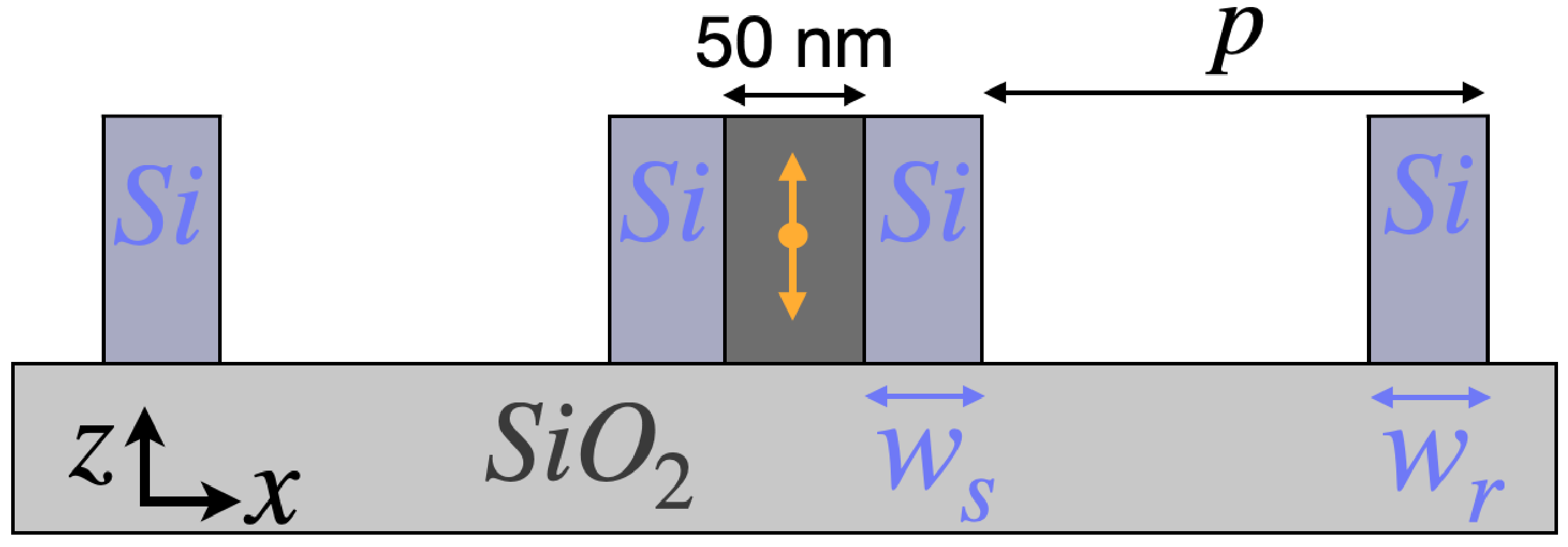}}
\includegraphics[width=0.5\columnwidth]{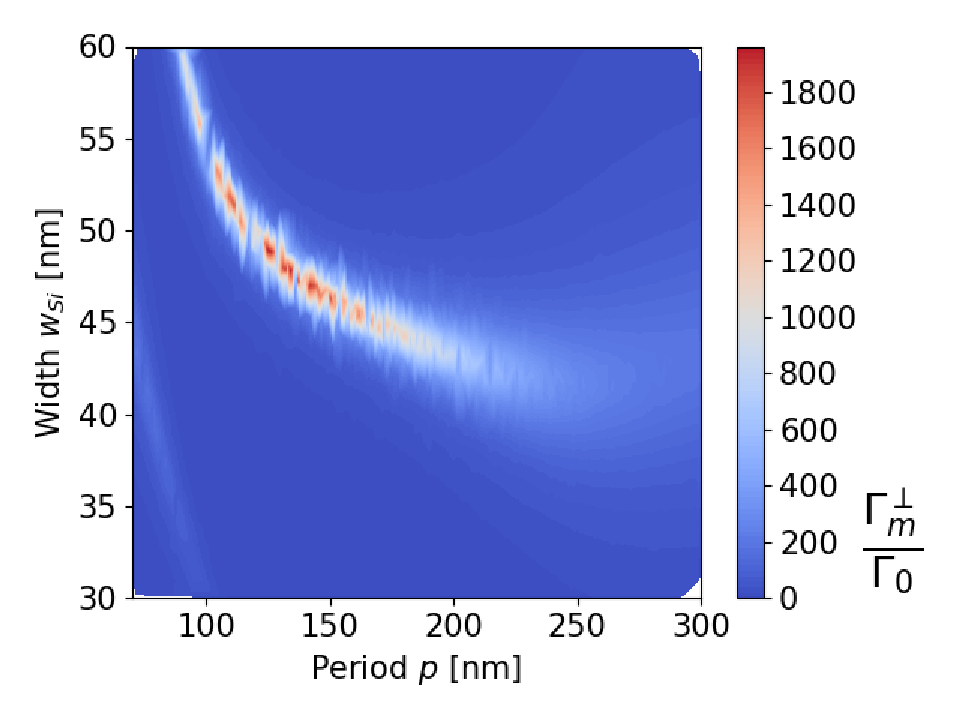}
\caption{Left : Scheme of the nanostructure to optimize. Right : Results of the Monte-Carlo optimization performed to optimize the width $w_{\text{Si}} = w_s = w_r$ and period $p$ dimensions of the grating to maximize the decay rate enhancement of a magnetic dipole located at the center of the structure and polarized perpendicularly to the substrate.}
\label{fig:OptMagZMC}
\end{figure}

The obtained optimized structures rely on Si core and a circular grating. Consequently, it presents resonances very sensitive to their shape. Thus, we complete the geometrical optimization by direct finite element method (FEM) simulations to further estimate the maximum achievable magnetic Purcell factor considering this configuration, schemed in Fig.~\ref{fig:OptMagZ}. 
We assume an out of plane MD located at the center of a Gd$_2$O$_3$:Eu$^{3+}$ nanodisk of 50 nm diameter and 100 nm height. It is surrounded by Si cylindrical shell and a circular gratings constituted of 3 rings. For simplicity, we assume a ring width $w_r$ identical to the Si shell width $w_s$ surrounding the Gd$_2$O$_3$:Eu$^{3+}$ core. We note $w_{\text{Si}} = w_s = w_r$. This parameter and the period $p$ of the circular grating has been optimized thanks to a Monte-Carlo simulations of 5000 iterations, see figure~\ref{fig:OptMagZMC}. 
We observe a maximum magnetic decay rate enhancement of $\Gmper \simeq 1940$ for a period $p = 125$ nm and a width $w_{\text{Si}} = 49$ nm. This once more demonstrates very high value of the achievable magnetic Purcell factor, largely above the enhancement obtained in bulk high index materials  $\Gmper =n^3 \simeq 62$ for Si. Without the circular grating, the Purcell factor is 125 only, demonstrating the important role of the grating. The rings strongly increase the decay rate to 754 (one ring), 1819 (two rings) and up to 1943 for 3 rings. We do not observe further increase with additionnal rings. The in-plane magnetic Purcell factor is $\Gmpar = 45$ for this structure. We also estimate the electric Purcell factor for the same configuration at $\lambda_e = 610$ nm and obtain a value of 3.8 for a randomly oriented emitter ($\Gepar = 5.4$ and $\Geper = 0.5$).\\

\begin{figure}[!h]
\centering
\includegraphics[width=0.45\columnwidth]{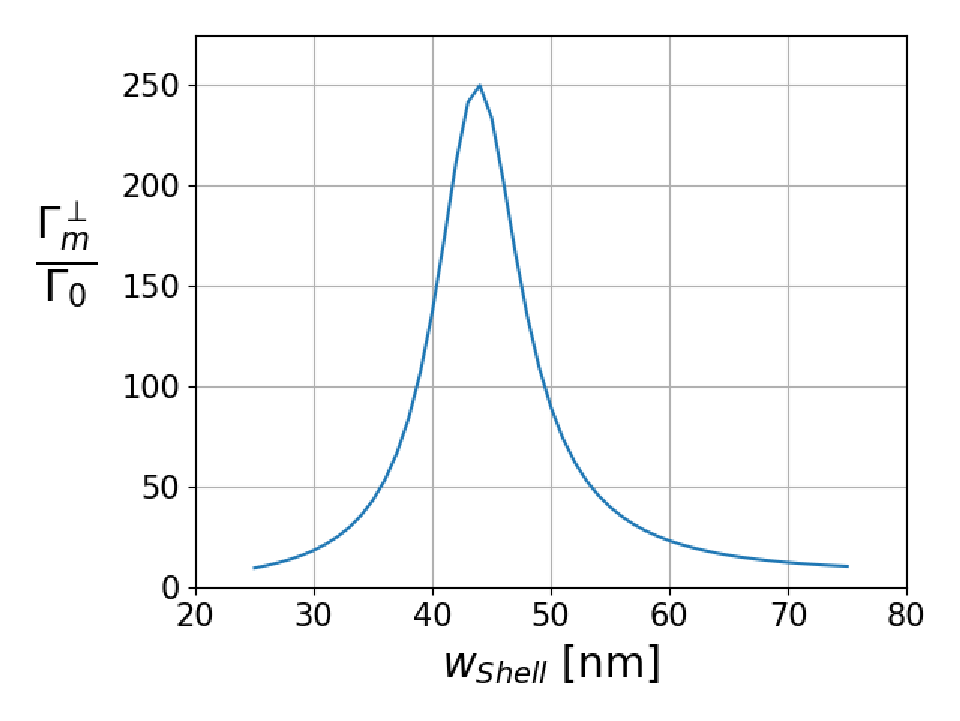}
\includegraphics[width=0.45\columnwidth]{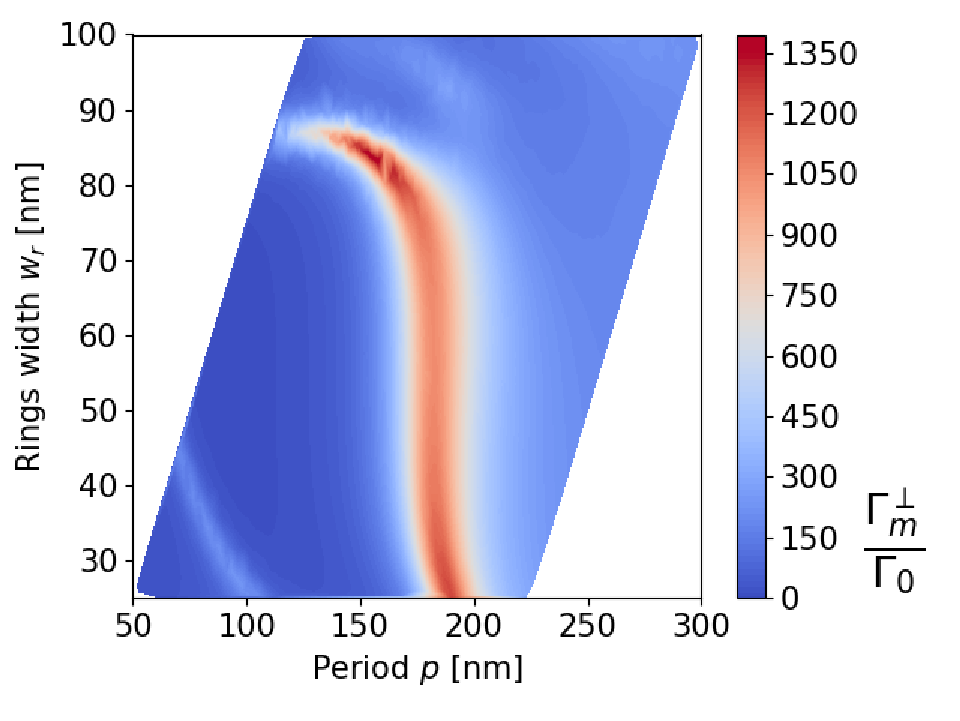}
\caption{Left: Magnetic Purcell factor $\Gmper$ dependency on the shell's width $w_{s}$ (no gratings). Right: Results of the Monte-Carlo with $w_{s} = 44$ nm to optimize the rings width $w_{r}$ and period $p$  dimensions of the grating to maximize the decay rate enhancement of a magnetic dipole located at the center of the structure and polarized perpendicularly to the substrate.}
\label{fig:OptMagZMC2step}
\end{figure}

If the Si cylinder shell (the core) is first optimized, without the circular grating, a maximum Purcell enhancement of 250 is achieved for $w_{s} = 44$ nm. Optimizing the rings width in a second step, we obtain $\Gmper \simeq 1390$ for a period $p = 157$ nm and $w_{r} = 83$ nm (see figure \ref{fig:OptMagZMC2step}). The three parameters, $w_s$, $w_r$ and $p$, should be optimized independently. However, we observe similar optimized parameters within the experimental fabrication precision so that specific numerical simulations would be performed after experimental characterization. We also observe on figure \ref{fig:OptMagZMC2step} a particular grating period ($p\simeq 185$ nm) for which the decay rate enhancement poorly depends on the width of the rings. This leads us to attribute the strong Purcell effect to the coupling to a defect mode into a photonic crystal. Indeed, considering the 1D equivalent multilayer Si/air configuration, we recognize the $\lambda/4$ Bragg mirror periodic arrangement A/B/A/B/... where A and B refers to Si and air layers respectively: $w_{air}=n_{Si}w_{Si}=\lambda_m/4$ = 147 nm leading to $w_{Si}$ = 37 nm and a period $p=w_{air}+w_{Si}=184$ nm. The cavity between the two Bragg mirrors corresponds approximately to the A/B/A/B/A/B/{\bf AA}/B/A/B/A/B/A arrangement that presents a defect mode at $\lambda_m=590$ nm \cite{Gaponenko:2008}. Here the circular grating plays a similar role, boosting the Purcell effect.
\\ 

\begin{figure}[!h]
\centering
\includegraphics[width=0.5\columnwidth]{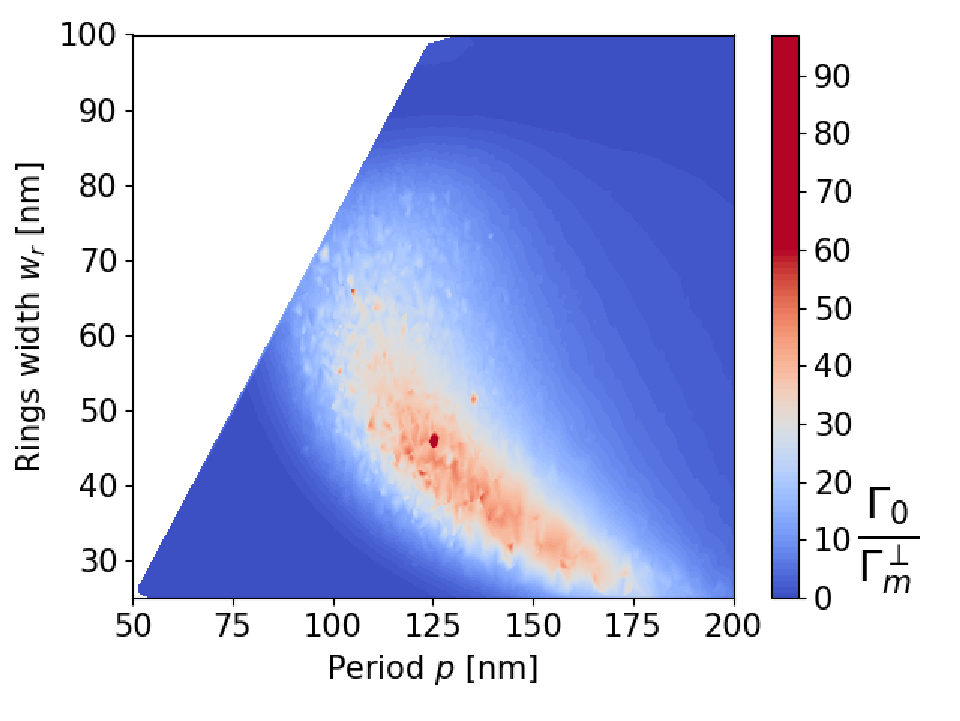}
\caption{Results of the Monte-Carlo optimization performed to optimize the rings width $w_r$ and period dimensions of a circular grating to minimize the decay rate of a out of plane MD located at the center of the structure. We plot the inverse of the Purcell factor to be maximized.}
\label{fig:OptMagZMCInhib}
\end{figure}

We have also investigated minimization of the out of plane magnetic Purcell factor performing a Monte-Carlo simulations with 5000 iterations by removing the Si shell around the Gd$_2$O$_3$:Eu$^{3+}$ core, see Fig. \ref{fig:OptMagZMCInhib}. The strongest inhibition is $\Gmper  \simeq 1/97$ obtained for a ring period $p = 125$ nm and width $w_r = 46$ nm. As expected, the geometry is complementary to the optimal design which leads to $\Gmper \simeq 1940$.\\

\section{Conclusions}
In summary, we have shown that evolutionary optimization and more precisely differential evolution algorithms are particularly relevant tools for the geometrical optimization of nanophotonics system. Its application to the design of dielectric planar nanoantennas for the enhancement and inhibition of the magnetic and electric Purcell factor has allowed us to retrieve regular and periodic characteristics that are naturally materialized through the evolution of these simulations. Regarding the geometry of the optimized nanoantennas' shape, the circular grating is a common pattern that is arising in the far-field region of the dipolar emission either of electric or magnetic nature. For the near-field region, different shapes have been retrieved and analyzed depending on the nature and orientation of the dipole such as dielectric nanodisks for MD or dielectric aperture nanoantennas for ED. We observe qualitative agreement with a deterministic approach based on the nanoscale design of the LDOS, revealing the underlying physical principles. With an emphasis on the magnetic emission, we also highlight the role of optical resonances thanks to modal analysis.  Completed with Monte-Carlo optimizations, we have been able to provide the design of an efficient planar Si nanoantenna that leads to a spontaneous magnetic decay rate enhancement of $\Gmper \simeq 2000$ in a Gd$_2$O$_3$:Eu$^{3+}$ core. The concrete realization of such an optimized nanostructure would lead to better control of the magnetic light-matter interaction for promising innovative applications in new types of light sources \cite{Staude:2019, Kalinic:2020}.

\section{Acknowledgment}
This work is supported by the French Investissements d'Avenir program EUR-EIPHI (17-EURE-0002), French National Research Agency (ANR) (project HiLight ANR-19-CE24-0026 and EQUIPEX+ SMARTLIGHT contract ANR-21-ESRE-0040), the European Regional Development Fund FEDER-FSE 2014-2020, Région Bourgogne- Franche-Comté.



\appendix
\newpage
\section{Detailed EO Results}
\label{sect:EOappendix}

\subsection{Out of plane magnetic dipole}
\begin{figure}[!h]
\centering
\includegraphics[width=0.49\columnwidth]{OptMagZ300V7.eps}
\includegraphics[width=0.49\columnwidth]{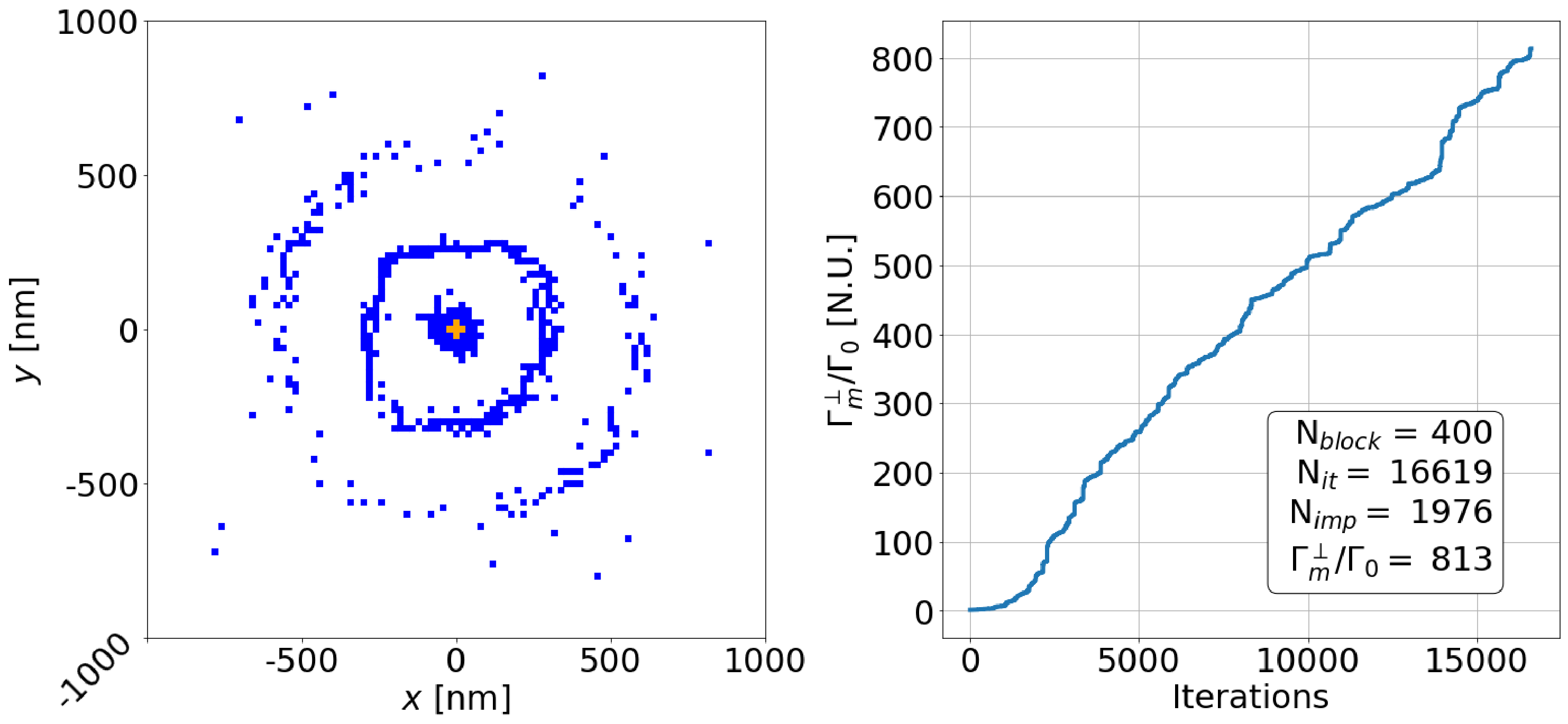}
\includegraphics[width=0.49\columnwidth]{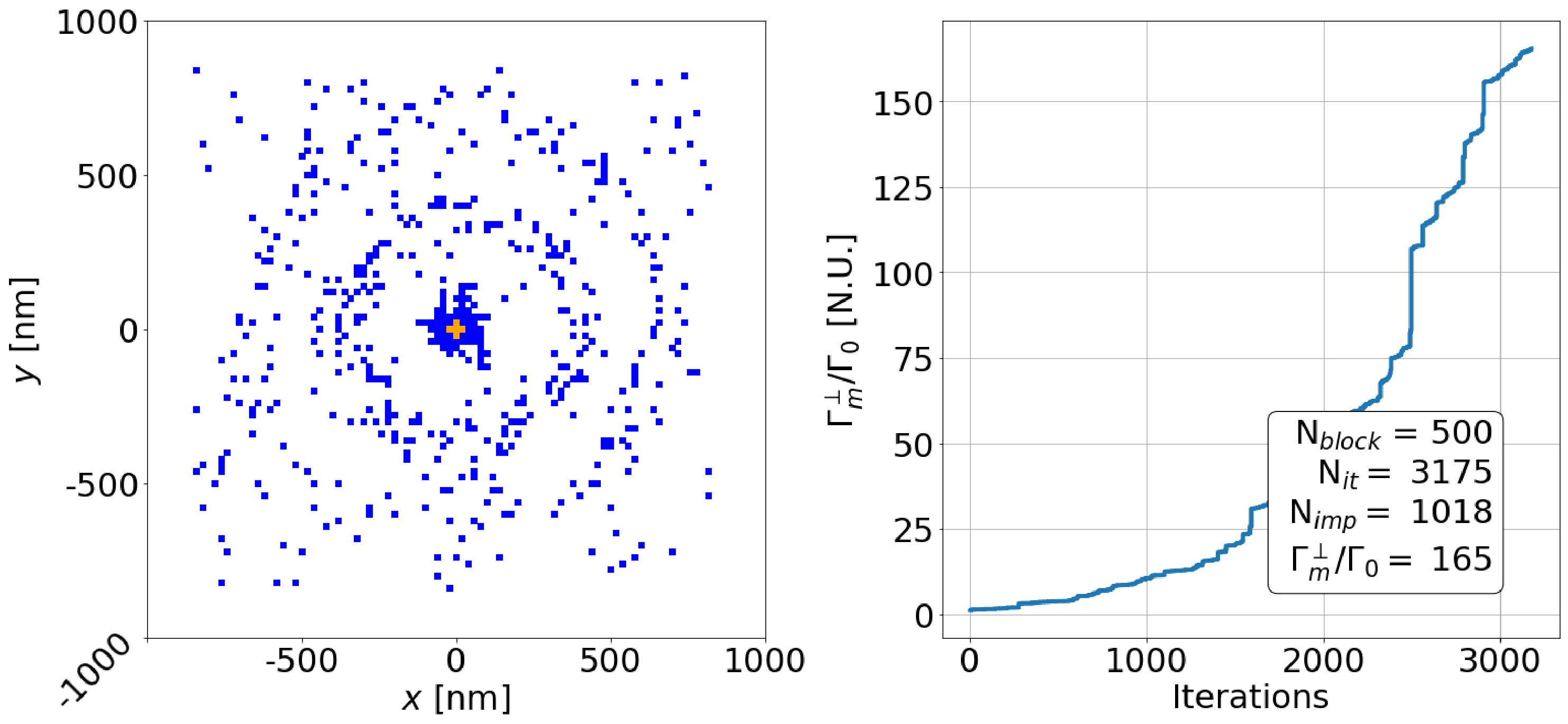}
\includegraphics[width=0.49\columnwidth]{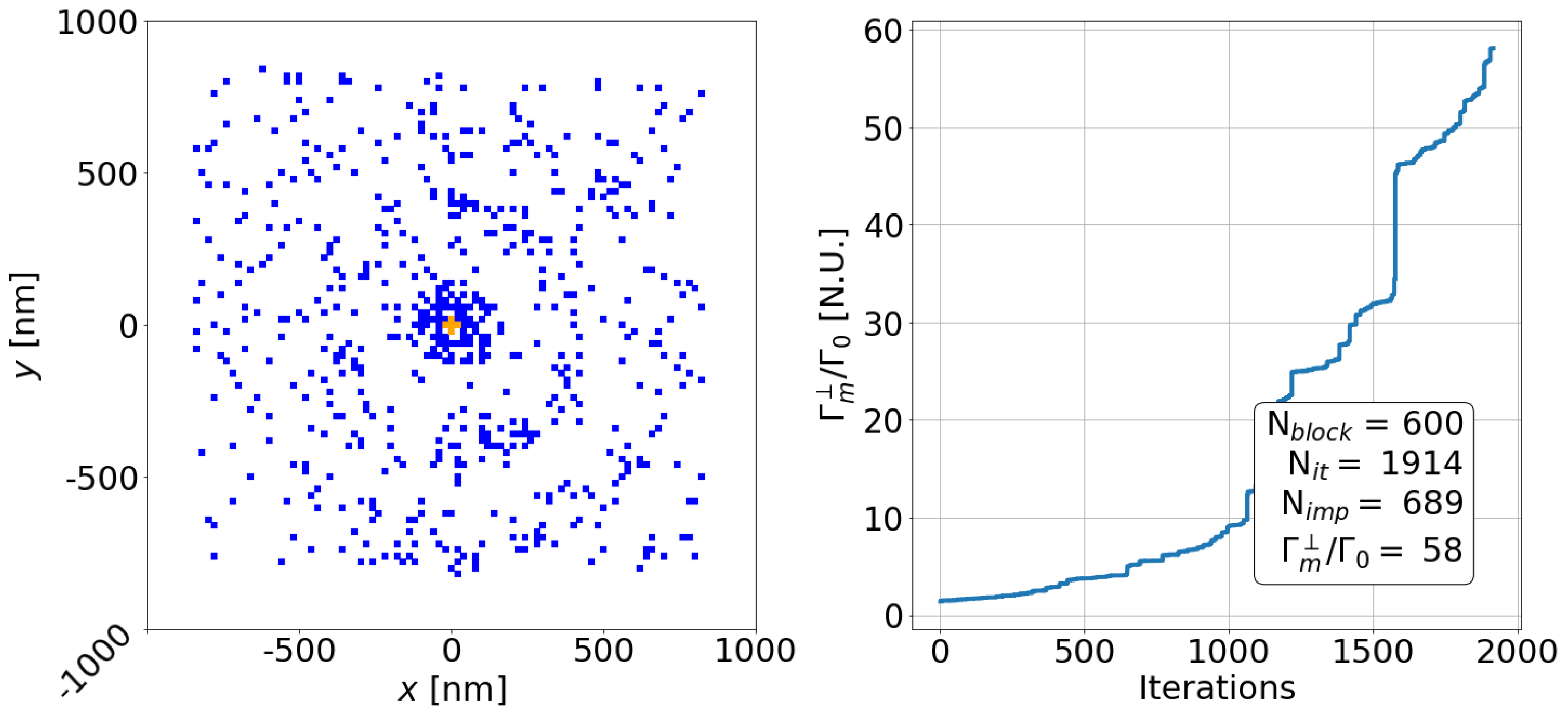}
\caption{Left : $XY$-plane projection of the optimized structure for $N = $ 300, 400, 500 and 600 Si nanopillars (orange : fixed core emitter, blue : Si nanopillars), Right : Evolution of the magnetic decay rate enhancement $\Gmper$ through the optimization iterations.}
\label{fig:OptMagZ3456}
\end{figure}

Figure~\ref{fig:OptMagZ3456} presents EO optimizations of the out of plane magnetic Purcell factor considering $N = $ 300, 400, 500 and 600 Si nanopillars.

\begin{figure}[!h]
\centering
\includegraphics[width=0.4\columnwidth]{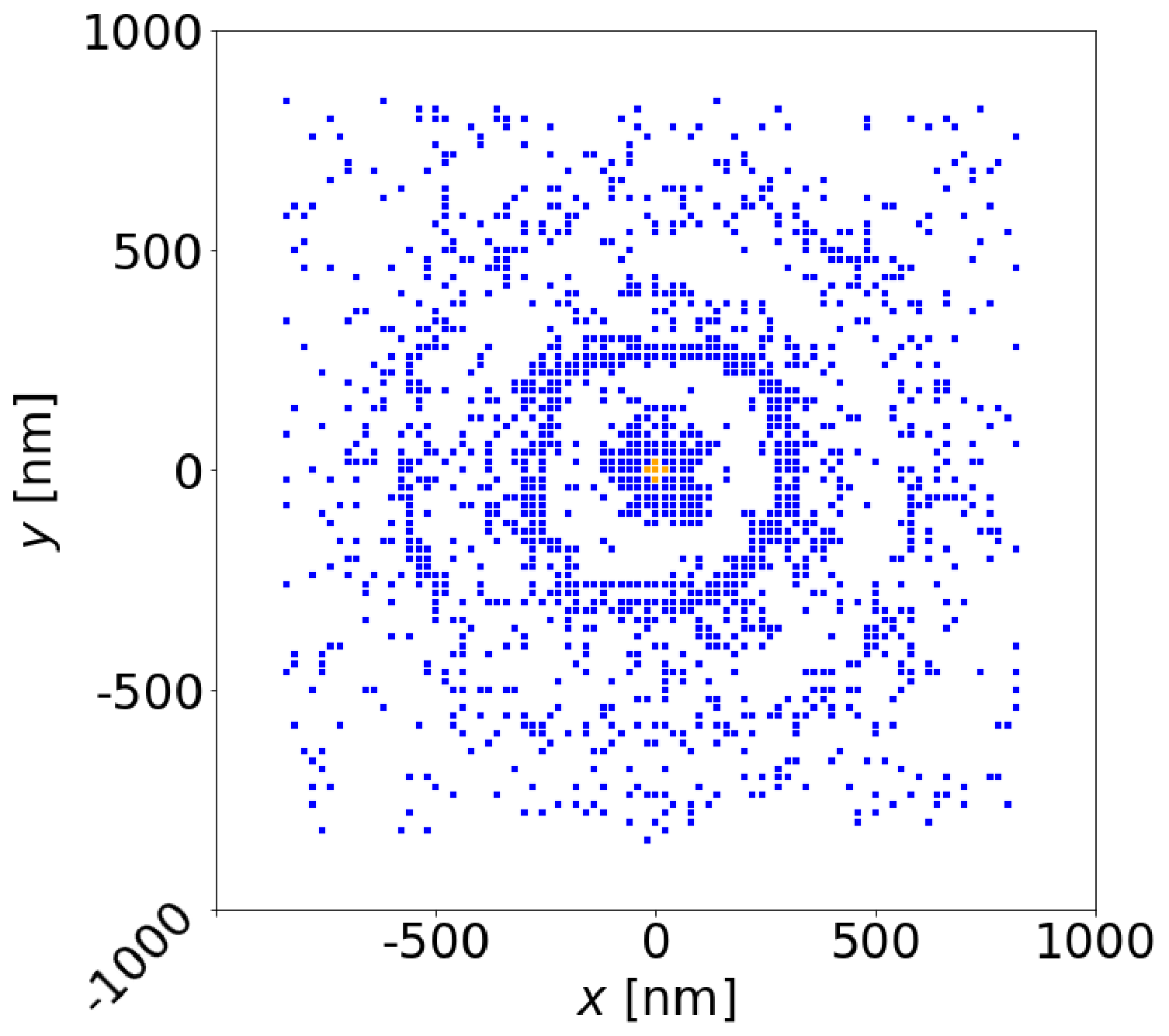}
\caption{$XY$-plane projection of the structure arising from the superposition of the four precedent ones (orange : fixed core emitter, blue : Si nanopillars).}
\label{fig:OptMagZSup}
\end{figure}

We also present in Fig.~\ref{fig:OptMagZSup} the superimposition of the four EO optimized structures. This reveals that all the optimizations converge towards a similar configuration, namely a Si core and a circular grating ring. 

\subsection{In plane magnetic dipole}
\begin{figure}[!h]
\centering
\includegraphics[width=0.49\columnwidth]{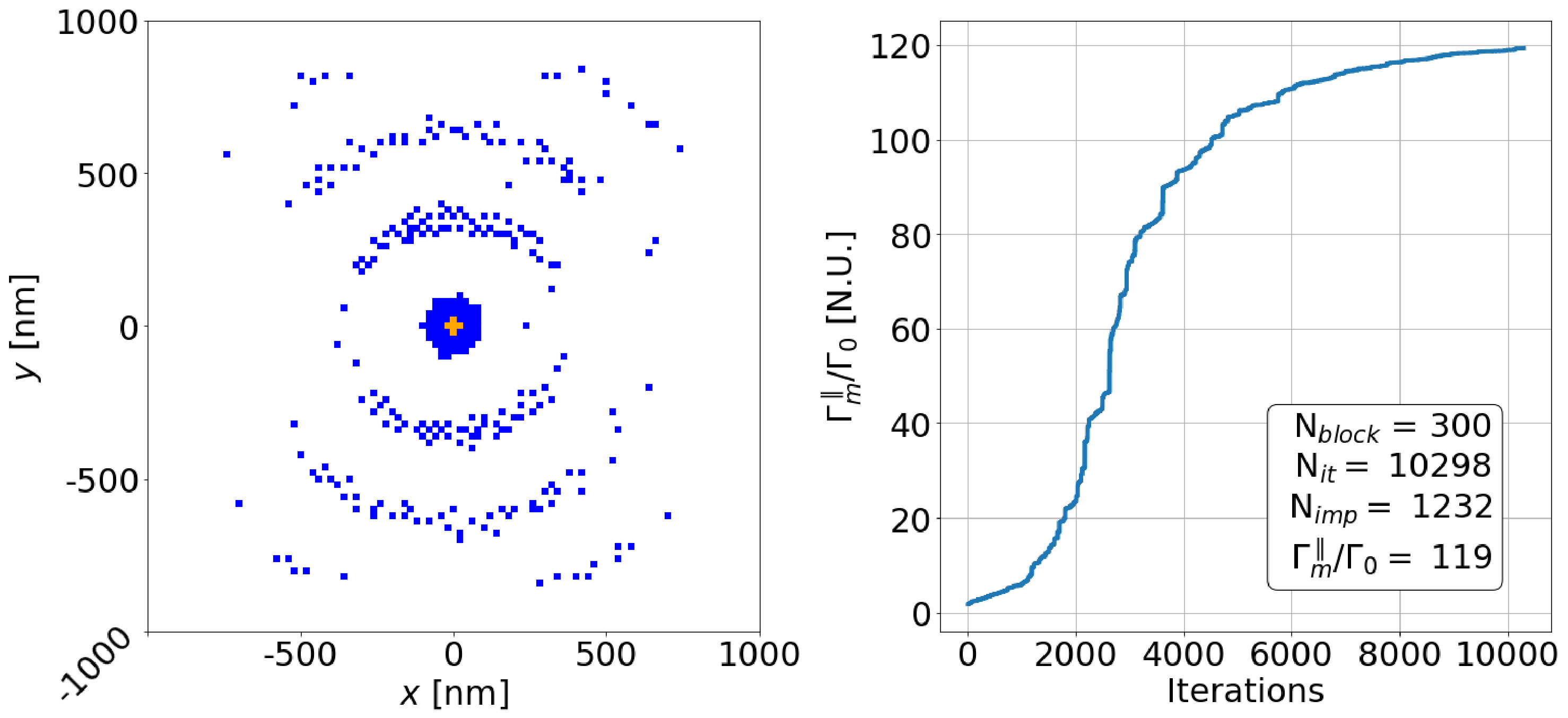}
\includegraphics[width=0.49\columnwidth]{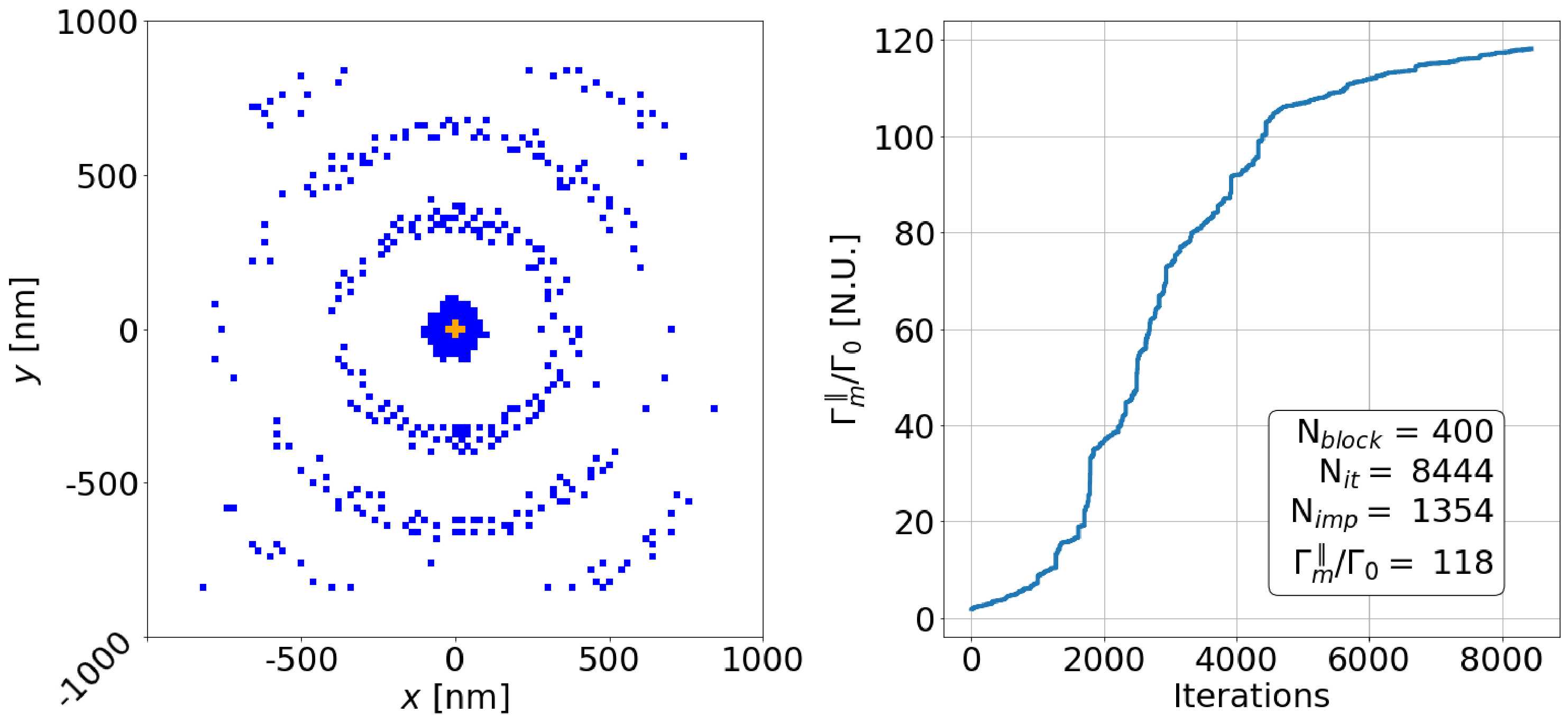}
\includegraphics[width=0.49\columnwidth]{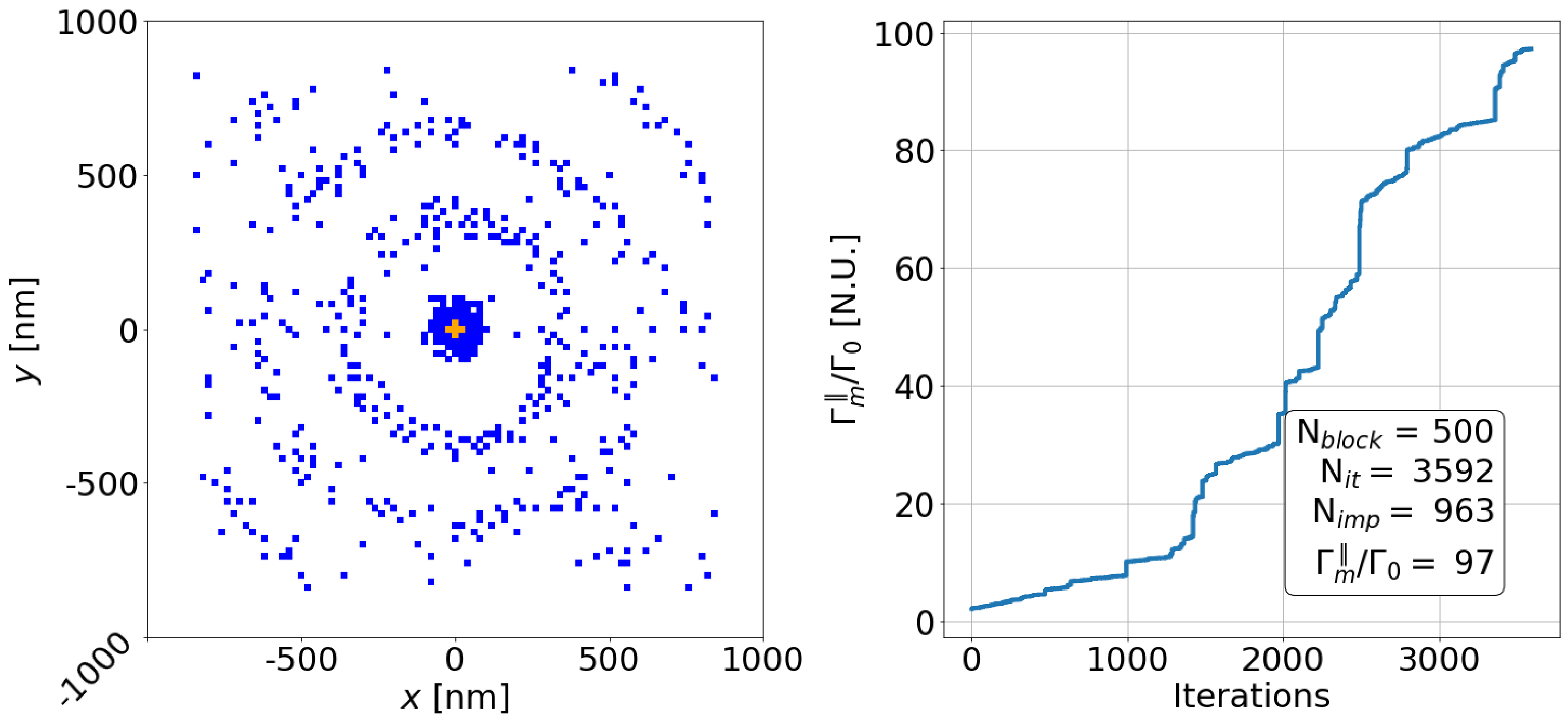}
\includegraphics[width=0.49\columnwidth]{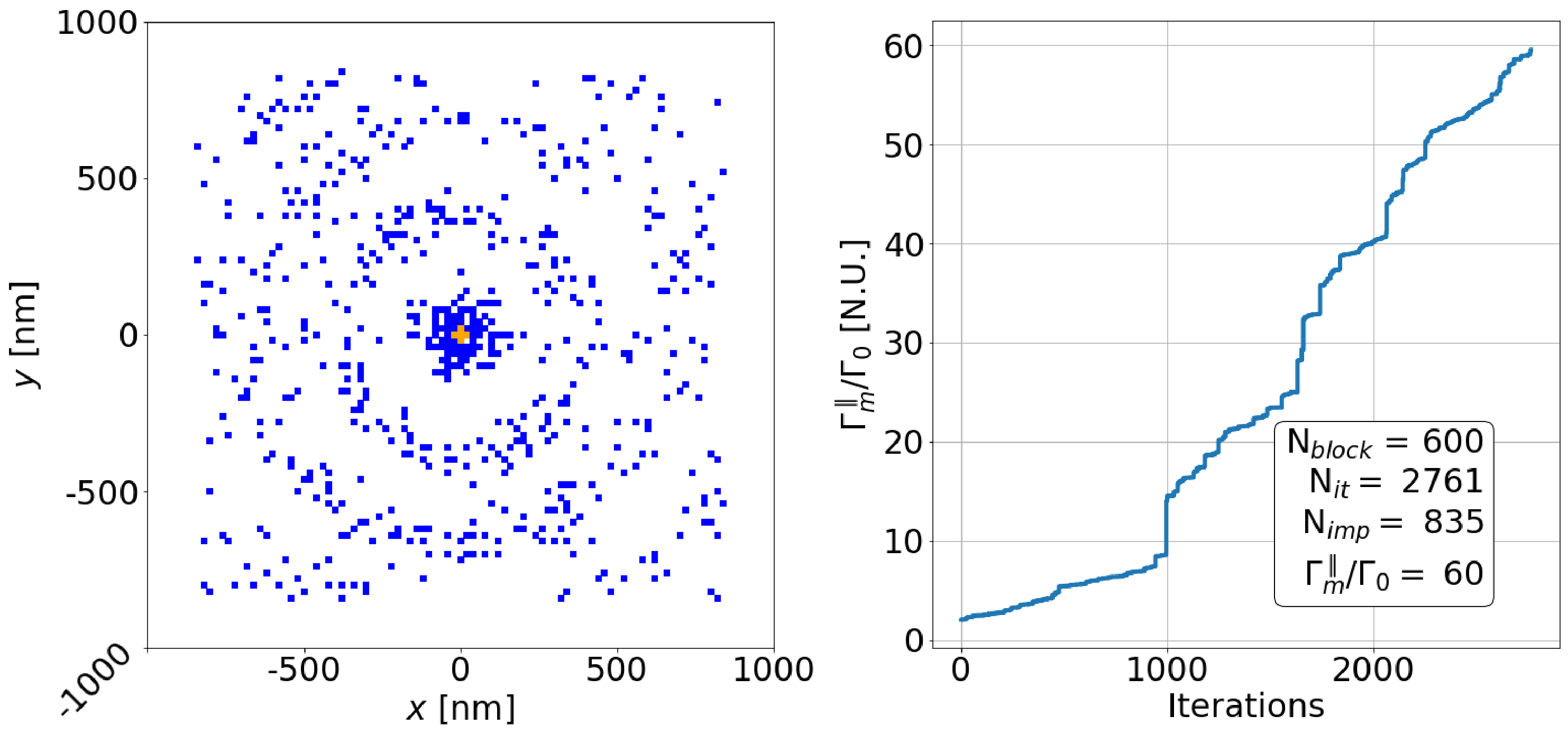}
\caption{Left : $XY$-plane projection of the optimized structure for $N = $ 300, 400, 500 and 600 Si nanopillars (orange : fixed core emitter, blue : Si nanopillars), Right : Evolution of the magnetic decay rate enhancement $\Gmpar$ through the optimization iterations. The MD is oriented along the $x$-axis.}
\label{fig:OptMagX3456}
\end{figure}

\begin{figure}[!h]
\centering
\includegraphics[width=0.4\columnwidth]{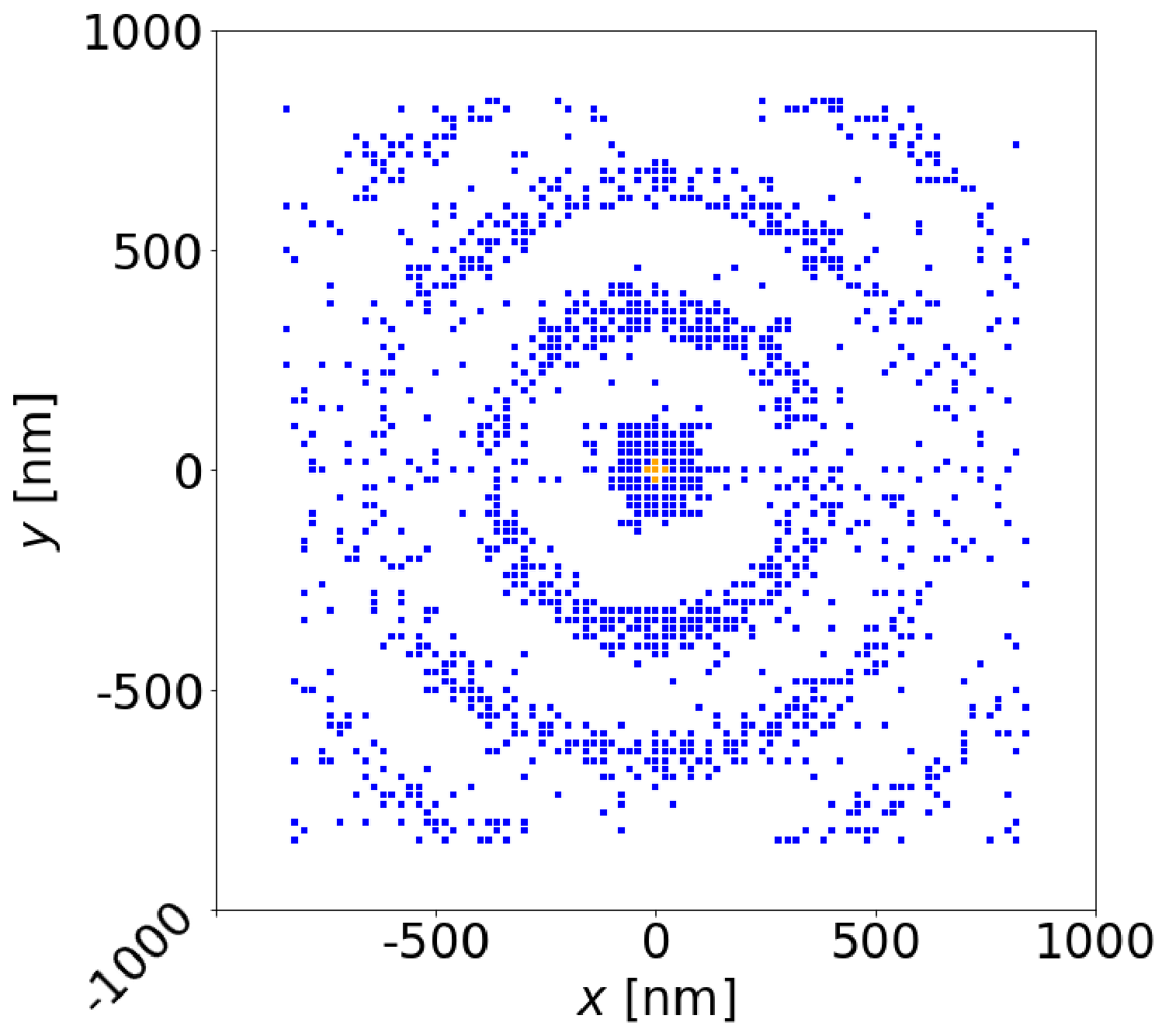}
\caption{$XY$-plane projection of the structure arising from the superposition of the four precedent ones (orange : fixed core emitter, blue : Si nanopillars).}
\label{fig:OptMagXSup}
\end{figure}

Figure~\ref{fig:OptMagX3456} presents optimizations of the in-plane MD for $N = $ 300, 400, 500 and 600 Si nanopillars. We obtain an optimized Purcell factor of about $\Gmpar \simeq 120$ for a Si core and grating. Once again, the simulation converges towards a regular and periodic structure very similar to the one obtained for the out of plane MD. However, the Si grating seems not fully circular but rather presents two lobes perpendicularly to the dipole orientation as for in-plane ED (see Fig. 4 in the main text). Again, the superimposition of the optimized structures, shown in Fig. \ref{fig:OptMagXSup}, reveals the similarity of the obtained structures. 

\subsection{Out of plane electric dipole}
\begin{figure}[!h]
\centering
\includegraphics[width=0.49\columnwidth]{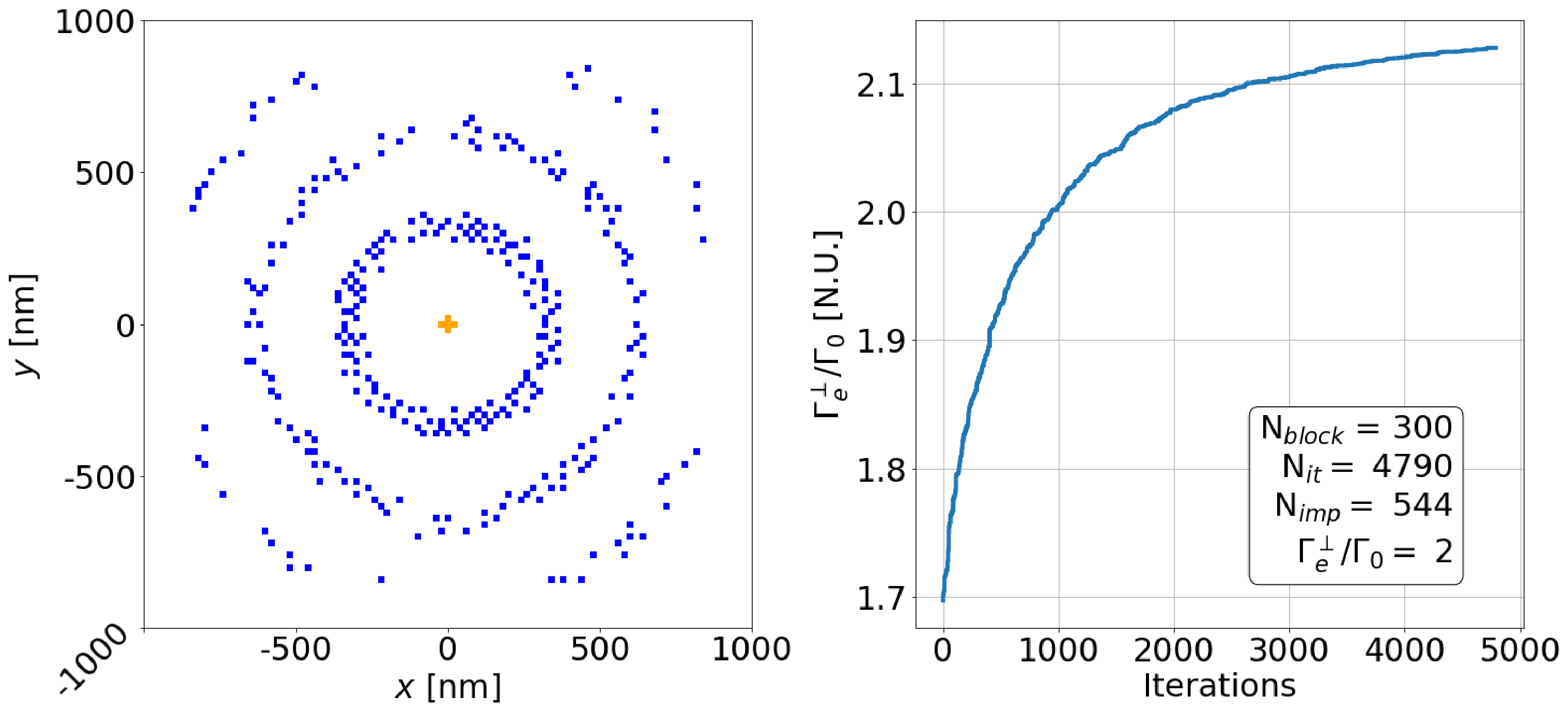}
\includegraphics[width=0.49\columnwidth]{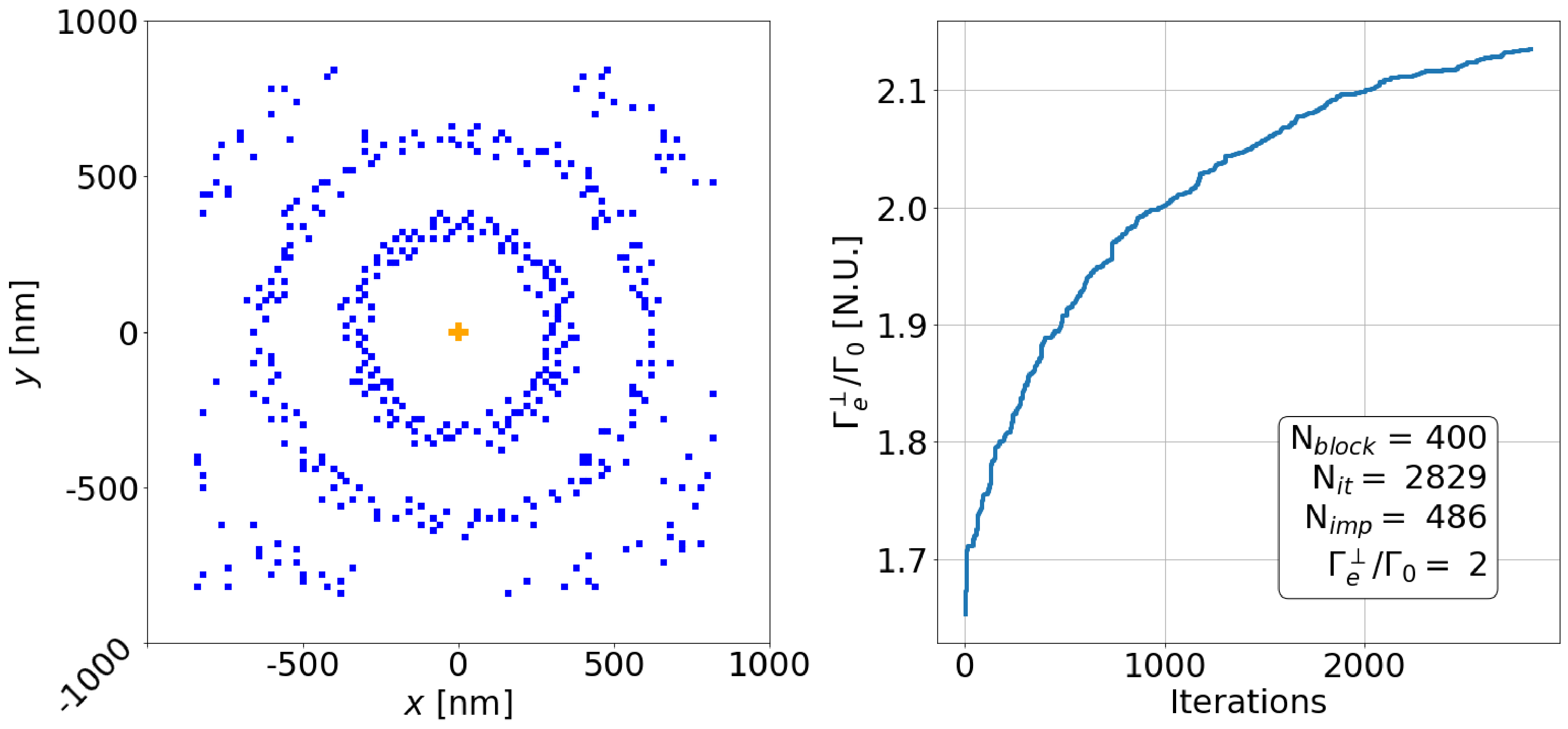}
\includegraphics[width=0.49\columnwidth]{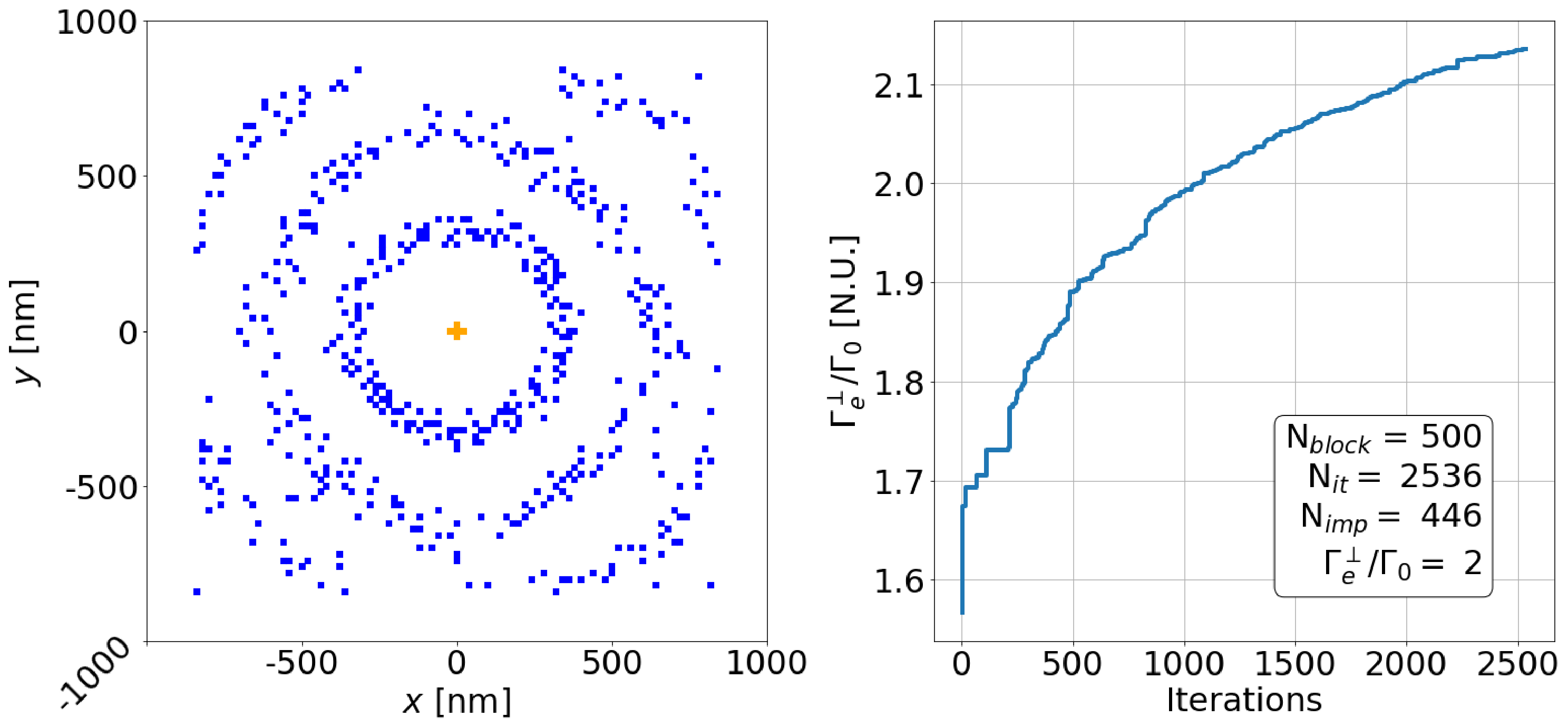}
\includegraphics[width=0.49\columnwidth]{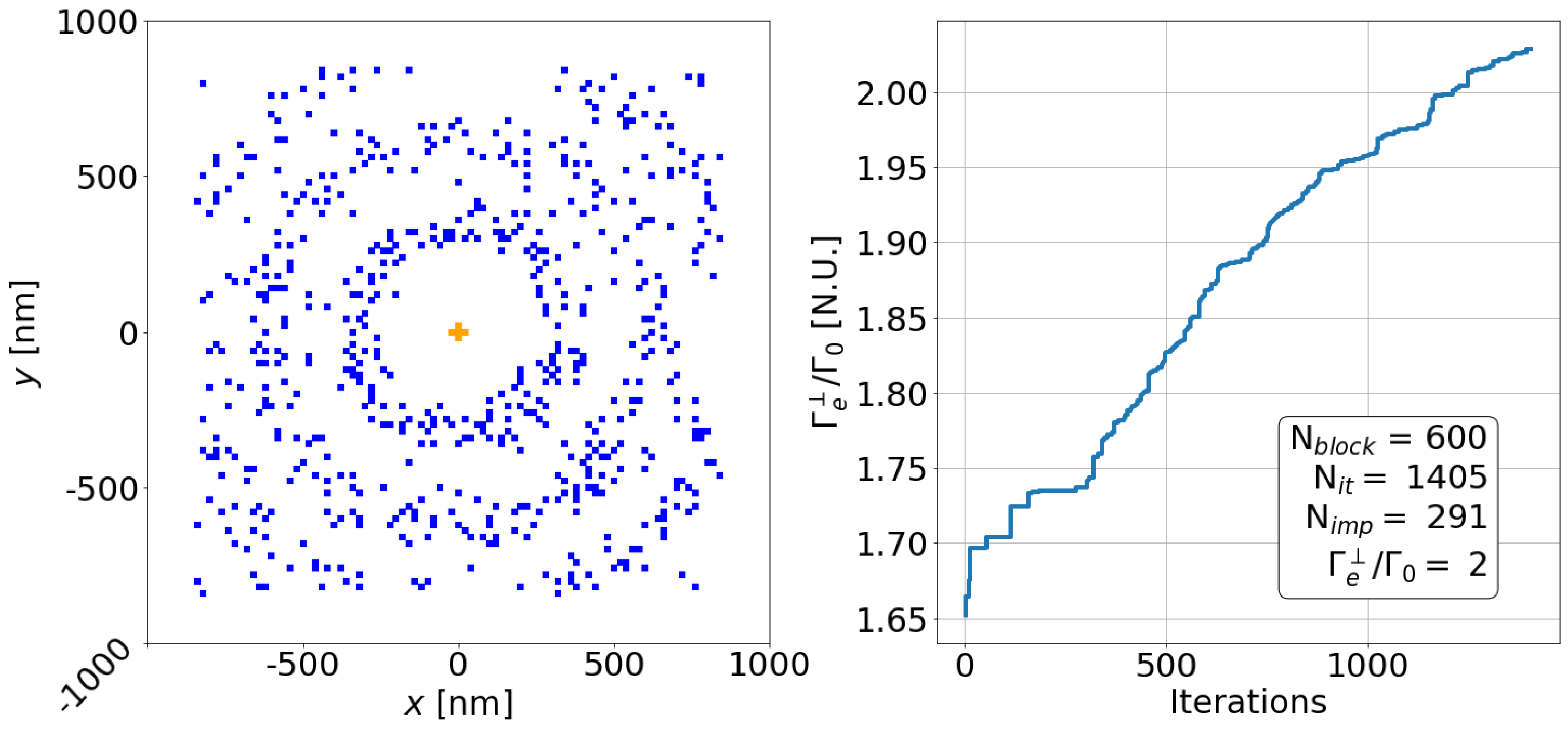}
\caption{Left : $XY$-plane projection of the optimized structure for $N = $ 300, 400, 500 and 600 Si nanopillars (orange : fixed core emitter, blue : Si nanopillars), Right : Evolution of the electric decay rate inhibition $\Geper$ through the optimization iterations.}
\label{fig:OptElecZ3456}
\end{figure}

\begin{figure}[!h]
\centering
\includegraphics[width=0.4\columnwidth]{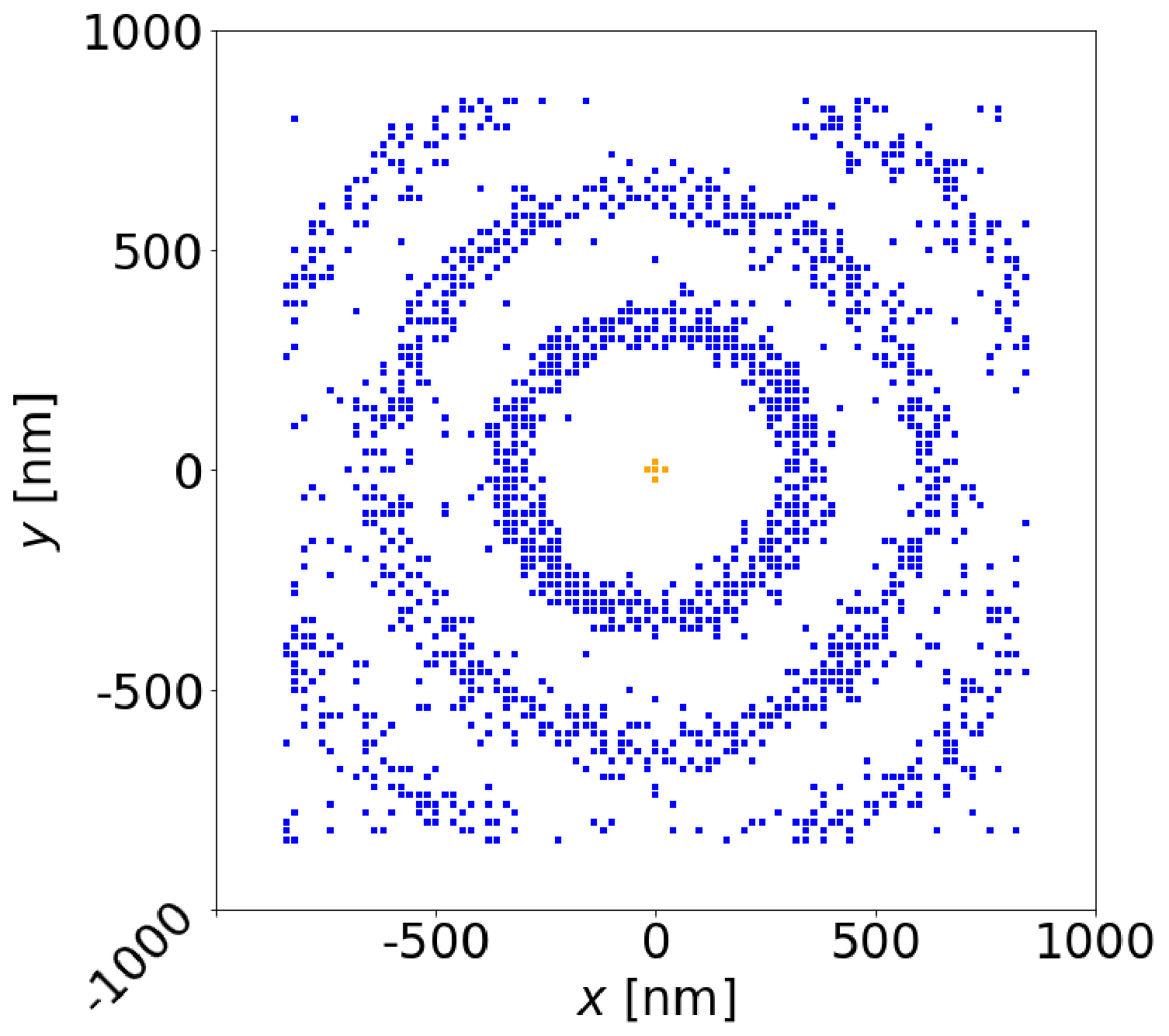}
\caption{$XY$-plane projection of the structure arising from the superposition of the four precedent ones (orange : fixed core emitter, blue : Si nanopillars).}
\label{fig:OptElecZSup}
\end{figure}

Figure~\ref{fig:OptElecZ3456} presents the results of the exaltation optimization for the electric dipolar emission at $\lambda_e$ and for a dipole polarized perpendicularly to the substrate for $N = $ 300, 400, 500 and 600 Si nanopillars.  
We observe very low enhancement of the Purcell factor ($\Geper=2$), achieved with Si circular grating but without any core. The superimposition of the optimized structures is shown in Fig. \ref{fig:OptElecZSup}, facilitating the comparison between the optimized structures, all presenting similar features. 

\subsection{In-plane electric dipole}
\begin{figure}[!h]
\centering
\includegraphics[width=0.49\columnwidth]{OptElecX300V6.eps}
\includegraphics[width=0.49\columnwidth]{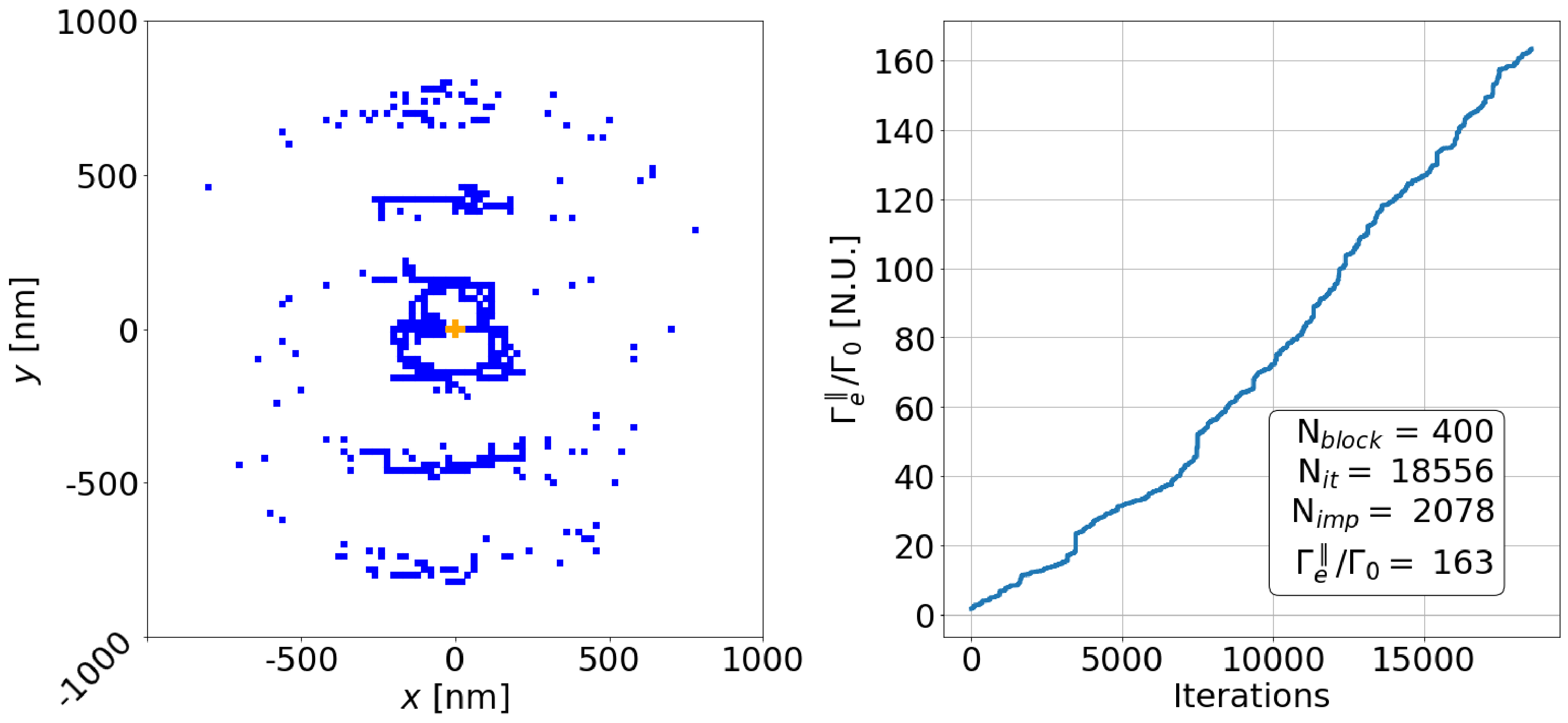}
\includegraphics[width=0.49\columnwidth]{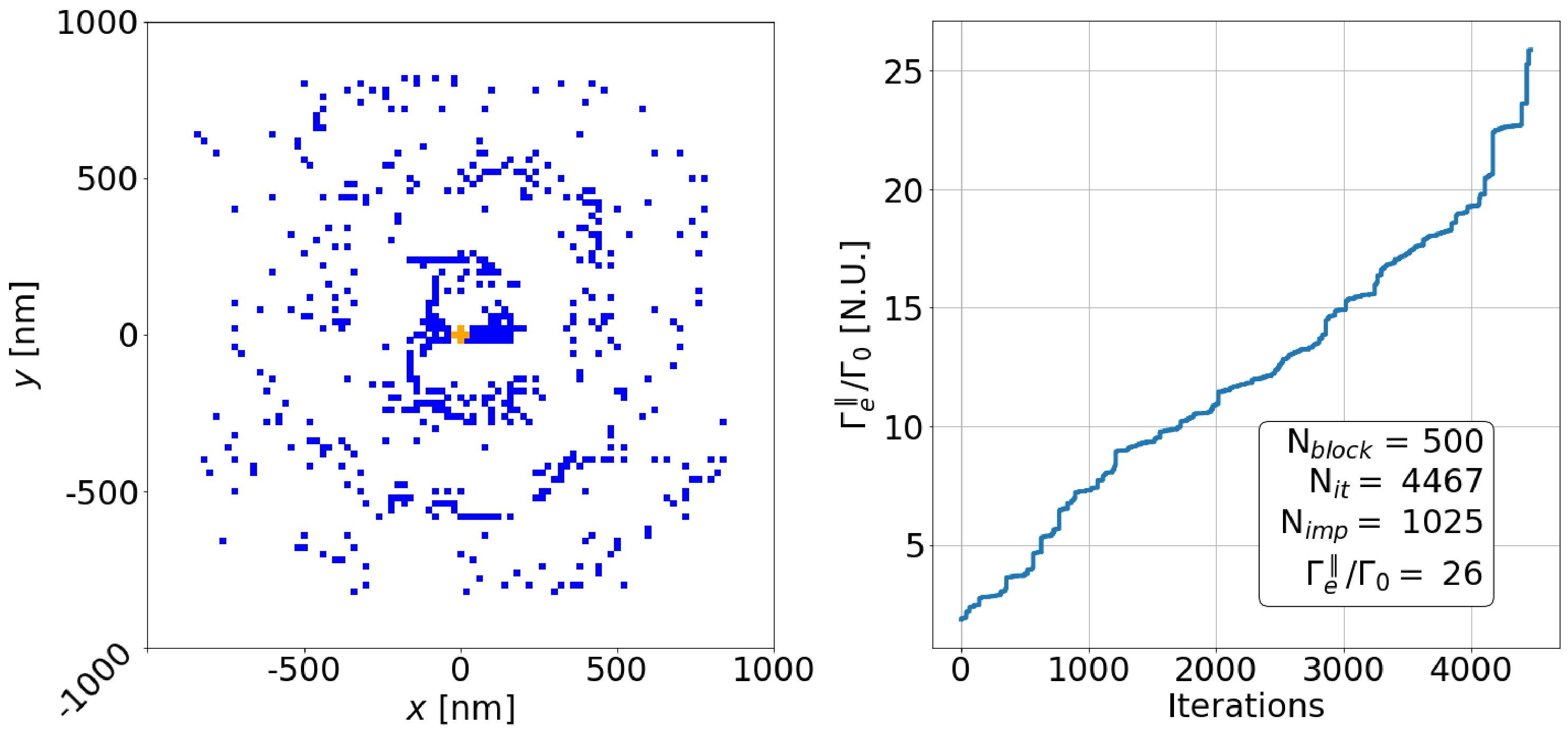}
\includegraphics[width=0.49\columnwidth]{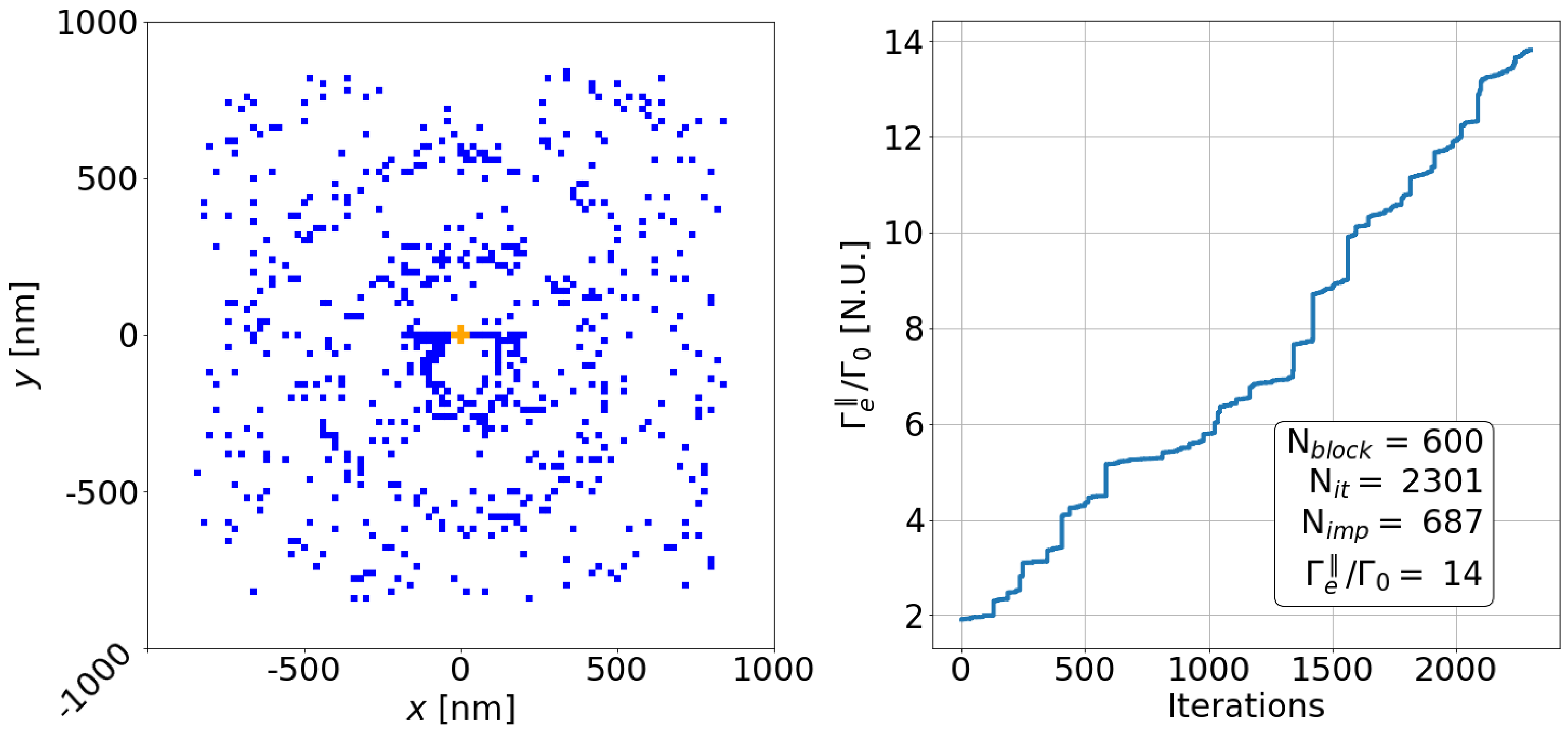}
\caption{Left : $XY$-plane projection of the optimized structure for $N = $ 300, 400, 500 and 600 Si nanopillars (orange : fixed core emitter, blue : Si nanopillars), Right : Evolution of the electric decay rate enhancement $\Gepar$ through the optimization iterations. The ED is oriented along the $x$-axis.}
\label{fig:OptElecX3456}
\end{figure}

Figure~\ref{fig:OptElecX3456} presents the results of the exaltation optimization for the in plane ED for $N = $ 300, 400, 500 and 600 Si nanopillars.
This leads to the superimposition presents on Fig. 5 and discuss in the main text.

\section{Local contribution to the electric or magnetic LDOS}
\label{sect:Mignuzzi}
Following the work of Mignuzzi and coworkers \cite{Mignuzzi:2019bis}, we derive the local contribution to the magnetic decay rate discussed in the main text (Eqs. (4-6)). 
Let us summarize their derivation that concerns an electric dipole before generalizing to the magnetic case. The decay rate associated to an electric dipole $\mathbf{d}$ is expressed as 
\begin{eqnarray}
\frac{\Gamma_e}{\Gamma_0}=1+\frac{6\pi\epsilon_0}{k_0^3\vert d\vert^2}Im\left\{ \mathbf{d} \cdot \mathbf{E}_s(\mathbf{r}_d)\right\}
\end{eqnarray}
where $\mathbf{E}_s(\mathbf{r}_0)$ is the dipolar electric field scattered at the position of  the dipole in its complex surroundings. In addition, the reciprocity theorem states that for current source density ${\mathbf J}_0$
\begin{eqnarray}
\int d^3{\mathbf r} ~ {\mathbf J}_0({\mathbf r}) \cdot {\mathbf E}_s({\mathbf r}) =  \int d^3{\mathbf r} ~ {\mathbf J}_s({\mathbf r}) \cdot {\mathbf E}_0({\mathbf r})
 \label{eq:rcpq}
\end{eqnarray}
where $\mathbf{E}_0(\mathbf{r})$ is the {\it free-space} electric field due to current source ${\mathbf J}_0$, $\mathbf{E}_s(\mathbf{r})$ the  electric field scattered in the {\it complex surroundings} and ${\mathbf J}_s({\mathbf r})=-i\omega \epsilon_0 (\epsilon_r-1) \mathbf{E}(\mathbf{r})$ the induced current in the complex surroundings ($\mathbf{E}=\mathbf{E}_0+\mathbf{E}_s)$.
Finally, for a point like electric dipole ${\mathbf J}_0({\mathbf r}) =-i\omega {\mathbf d} \delta({\mathbf r}-{\mathbf r}_0)$ it comes

\begin{eqnarray}
 {\mathbf d}  \cdot {\mathbf E}_s({\mathbf r}_0) &=& \epsilon_0 \int d^3{\mathbf r} ~ \left(\epsilon_r({\mathbf r})-1\right){\mathbf E}_0({\mathbf r}) \cdot {\mathbf E}({\mathbf r})
\\
&=& \frac{1}{\epsilon_0} \int d^3{\mathbf r} ~ \left(\epsilon_r({\mathbf r})-1\right) \mathbf{d}\cdot {\mathbf G}_0({\mathbf r}_0,{\mathbf r}) \cdot {\mathbf G}({\mathbf r},{\mathbf r}_0) \cdot \mathbf{d}
\end{eqnarray}
where we introduced the free-space electric Green's tensor  ${\mathbf G}_0$ and the Green's tensor associated to the complex environment ${\mathbf G}$.
The electric decay rate modification follows
\begin{eqnarray}
&& \frac{\Gamma_e}{\Gamma_0}=1+\frac{6\pi}{k_0^3\vert d\vert^2}\int d^3{\mathbf r} ~ \left(\epsilon_r({\mathbf r})-1\right) Im\left[ f_E(\mathbf{r})\right] \\
&& f_E(\mathbf{r})=\mathbf{d} \cdot {\mathbf G}_0({\mathbf r}_0,{\mathbf r}) \cdot {\mathbf G}({\mathbf r},{\mathbf r}_0) \cdot \mathbf{d}
\end{eqnarray}

In case of a magnetic dipole emission  ${\mathbf m}$, the next term in the expansion of $\int d^3{\mathbf r} ~ {\mathbf J}_0({\mathbf r})$ in Eq. (\ref{eq:rcpq}) leads to \cite{Landau-Lifshitz:1960c}
\begin{eqnarray}
i\omega  {\mathbf m} \cdot {\mathbf B}_s({\mathbf r}_0) &=&  \int d^3{\mathbf r} ~ {\mathbf J}_s({\mathbf r}) \cdot {\mathbf E}_0({\mathbf r}) \\
 {\mathbf m} \cdot {\mathbf B}_s({\mathbf r}_0) &=&-\epsilon_0 \int d^3{\mathbf r} ~ \left(\epsilon_r({\mathbf r})-1\right) \mathbf{m}\cdot {\mathbf G}^{EH}_0({\mathbf r}_0,{\mathbf r}) \cdot {\mathbf G}^{EH}({\mathbf r},{\mathbf r}_0) \cdot \mathbf{m} \\
  &=&\mu_0 \int d^3{\mathbf r} ~ \left(\epsilon_r({\mathbf r})-1\right) \mathbf{m}\cdot {\mathbf G}^{HE}_0({\mathbf r}_0,{\mathbf r}) \cdot {\mathbf G}^{EH}({\mathbf r},{\mathbf r}_0) \cdot \mathbf{m}
\end{eqnarray}
where  the mixed Green's tensor ${\mathbf G}^{EH}$ (resp. ${\mathbf G}^{EH}_0$)  gives the electric field scattered by a magnetic dipole in the complex environment (resp. in free--space). Last line is obtained using ${\mathbf G}^{EH}_0=-(Z_0/\epsilon_0 c){\mathbf G}^{HE}_0=-(\mu_0/\epsilon_0){\mathbf G}^{HE}_0$.  Finally, The magnetic decay rate modification follows (see also Eq. 10 in ref. \cite{Wiecha:2018bis})
\begin{eqnarray}
 \frac{\Gamma_m}{\Gamma_0}&=&1+\frac{6\pi}{\mu_0k^3\vert m\vert^2}Im\left\{ \mathbf{m} \cdot \mathbf{B}_s(\mathbf{r}_d)\right\}\\
 &=&1+\frac{6\pi}{k^3\vert m\vert^2}\int d^3{\mathbf r} ~ \left(\epsilon_r({\mathbf r})-1\right) Im\left[ f_H(\mathbf{r})\right] \\
&& f_H(\mathbf{r})=\mathbf{m} \cdot {\mathbf G}_0^{HE}({\mathbf r}_0,{\mathbf r}) \cdot {\mathbf G}^{EH}({\mathbf r},{\mathbf r}_0) \cdot \mathbf{m}
\end{eqnarray}
It is worth noticing that this expression is consistent with the equation used for computing the magnetic decay rate \cite{Wiecha:2018bis,Majorel:2020bis}:
\begin{eqnarray}
&&\int d^3{\mathbf r} ~ \left(\epsilon_r({\mathbf r})-1\right) {\mathbf G}_0^{HE}({\mathbf r}_0,{\mathbf r}) \cdot {\mathbf G}^{EH}({\mathbf r},{\mathbf r}_0)=k_0^2{\mathbf G}^{HH}_s({\mathbf r}_0,{\mathbf r}_0) \;, \text{so ~that} \\
&&\frac{\Gamma_m}{\Gamma_0}=1+\frac{6\pi}{k_0\vert m\vert^2}Im \left[\mathbf{m} \cdot {\mathbf G}^{HH}_s({\mathbf r}_0,{\mathbf r}_0) \cdot \mathbf{m} \right]
\end{eqnarray}

\newcounter{pointnumberO}
\setcounter{pointnumberO}{0}
\addtocounter{pointnumberO}{1}
\makeatletter
\renewcommand{\@biblabel}[1]{#1.}
\makeatother

\end{document}